\documentclass[10pt,conference]{IEEEtran}
\IEEEoverridecommandlockouts		

\usepackage[ruled,vlined,lined,commentsnumbered]{algorithm2e}
\usepackage{booktabs}
\usepackage{moreverb}
\usepackage{fontenc}
\usepackage{amsmath}
\usepackage{amssymb}
\usepackage{fancybox}
\usepackage{color}
\usepackage{colortbl}
\usepackage{array}
\usepackage{multirow}
\usepackage{multicol}
\usepackage{listings}
\usepackage{makecell}
\usepackage{graphicx}
\usepackage{setspace}
\usepackage{soul}
\usepackage{siunitx}

\usepackage{balance}
\usepackage{csvsimple}
\usepackage{longtable}
\usepackage[skip=0pt,labelfont=bf]{caption}
\usepackage{url}
\usepackage{lscape}
\usepackage{rotating}
\usepackage{tikz}
\usepackage{enumitem}
\usepackage{subcaption}

\usepackage[most]{tcolorbox}

\usepackage{threeparttable}

\usepackage{bclogo}

\usepackage{marvosym}
\usepackage{pifont}% http://ctan.org/pkg/pifont%\usepackage{subfigure}

\usepackage{calligra}
\usepackage{algorithmicx}
\usepackage{algpseudocode}
\usepackage{tabularx}

\algnewcommand\algorithmicforeach{\textbf{for each}}
\algdef{S}[FOR]{ForEach}[1]{\algorithmicforeach\ #1\ \algorithmicdo}

\newcolumntype{L}[1]{>{\raggedright\let\newline\\\arraybackslash\hspace{0pt}}m{#1}}
\newcolumntype{C}[1]{>{\centering\let\newline\\\arraybackslash\hspace{0pt}}m{#1}}
\newcolumntype{R}[1]{>{\raggedleft\let\newline\\\arraybackslash\hspace{0pt}}m{#1}}

\definecolor{codegreen}{rgb}{0,0.6,0}
\definecolor{codegray}{rgb}{0.5,0.5,0.5}
\definecolor{codepurple}{rgb}{0.58,0,0.82}
\definecolor{backcolour}{rgb}{0.95,0.95,0.92}
\definecolor{lightgray}{gray}{0.9}

\lstdefinestyle{mystyle}{
    commentstyle=\color{codegreen},
    keywordstyle=\color{magenta},
    numberstyle=\tiny\color{codegray},
    stringstyle=\color{codepurple},
    basicstyle=\footnotesize,
    breakatwhitespace=false,
    breaklines=true,
    captionpos=b,
    keepspaces=true,
    showspaces=false,
    showstringspaces=false,
    showtabs=false,
    tabsize=2
}

\lstdefinelanguage{diff}{
  morecomment=[f][\color{blue}]{@@},     % group identifier
  morecomment=[f][\color{red}]-,         % deleted lines
  morecomment=[f][\color{codegreen}]+,       % added lines
  morecomment=[f][\color{red}]{---}, % Diff header lines (must appear after +,-)
  morecomment=[f][\color{codegreen}]{+++},
}

\setlist{noitemsep} %to leave space around whole list

\definecolor{darkpastelred}{rgb}{0.76, 0.23, 0.13}
\definecolor{ao(english)}{rgb}{0.0, 0.5, 0.0}

\definecolor{darkpastelred}{rgb}{0.76, 0.23, 0.13}
\definecolor{ao(english)}{rgb}{0.0, 0.5, 0.0}

\lstdefinelanguage{diff}{
  morecomment=[f][\color{blue}]{@@},     % group identifier
  morecomment=[f][\color{red}]-,         % deleted lines
  morecomment=[f][\color{codegreen}]+,       % added lines
  morecomment=[f][\color{red}]{---}, % Diff header lines (must appear after +,-)
  morecomment=[f][\color{codegreen}]{+++},
}

\definecolor{yellow}{RGB}{255,255,153}
\definecolor{grey}{RGB}{224,224,224}

\newboolean{showcomments}
\setboolean{showcomments}{true}
\ifthenelse{\boolean{showcomments}}
 { \newcommand{\mynote}[2]{
      \fbox{\bfseries\sffamily\scriptsize#1}
        {\small$\blacktriangleright$\textsf{\emph{#2}}$\blacktriangleleft$}}}
        { \newcommand{\mynote}[2]{}}

\definecolor{DarkOrange}{rgb}{0.8,0.3,0.0}
\definecolor{DarkCyan}{rgb}{0.0, 0.55, 0.55}
\definecolor{DarkCyel}{rgb}{1.0, 0.49, 0.0}
\definecolor{yellow-green}{rgb}{0.6, 0.8, 0.2}

\newcolumntype{?}{!{\vrule width 1pt}}

\newcommand{\toolname}{{\sc FlexiRepair}\xspace}

\newcommand{\miner}{{\sc Miner}\xspace}
\newcommand{\inferrer}{{\sc Inferrer}\xspace}
\newcommand{\generator}{{\sc Generator}\xspace}
\newcommand{\spinfer}{{\sc Spinfer}\xspace}

\newcommand{\note}[1]{
\begin{tcolorbox}[tile,size=fbox,boxsep=1.5mm,boxrule=0pt,top=0pt,bottom=0pt,
borderline west={1mm}{-2pt}{black!50!white},colback=black!5!white]
\em #1
\end{tcolorbox}
}

\usepackage[nounderscore]{syntax}
\usepackage{newfloat}

\DeclareFloatingEnvironment[
  fileext   = logr,% don't use log here!
  listname  = {List of Grammars},
  name      = Grammar,
  placement = htp
]{Grammar}

\begin{document}
\title{{\sc FlexiRepair}: 
%a Flexible, Transparent and Practical Pipeline for Template-based Program Repair
%Building on Semantic Patches for Program Repair    
Transparent Program Repair with Generic Patches
}

\author{Anil Koyuncu, Tegawend\'e F. Bissyand\'e, Jacques Klein,
 Yves Le Traon \\\textit{SnT, University of Luxembourg - Luxembourg}}

% \IEEEauthorblockA{\textit{SnT, University of Luxembourg - Luxembourg} \\}
% \affiliation{%
%   \institution{SnT, University of Luxembourg - Luxembourg}
% \settopmatter{printacmref=false}

%\begin{CCSXML}
%<ccs2012>
%<concept>
%<concept_id>10011007.10011074.10011099</concept_id>
%<concept_desc>Software and its engineering~Software verification and validation</concept_desc>
%<concept_significance>500</concept_significance>
%</concept>
%<concept>
%<concept_id>10011007.10011074.10011099.10011102</concept_id>
%<concept_desc>Software and its engineering~Software defect analysis</concept_desc>
%<concept_significance>300</concept_significance>
%</concept>
%<concept>
%<concept_id>10011007.10011074.10011099.10011102.10011103</concept_id>
%<concept_desc>Software and its engineering~Software testing and debugging</concept_desc>
%<concept_significance>100</concept_significance>
%</concept>
%</ccs2012>
%\end{CCSXML}
%
%\ccsdesc[500]{Software and its engineering~Software verification and validation}
%\ccsdesc[300]{Software and its engineering~Software defect analysis}
%\ccsdesc[100]{Software and its engineering~Software testing and debugging}
%
\maketitle

\begin{abstract}
Template-based program repair research is in need for a common ground to express fix patterns in a standard and reusable manner. We propose to build on the concept of {\em generic patch} (also known as semantic patch), which is widely used in the Linux community to automate code evolution. We advocate that generic patches could provide at the same time a unified representation and a specification for fix patterns. Generic patches are indeed formally defined, and there exists a robust, industry-adapted, and extensible engine that processes generic patches to perform control-flow code matching and automatically generates concretes patches based on the specified change operations.

In this paper, we present the design and implementation of a repair framework, \toolname, that explores generic patches as the core concept. In particular, we show how concretely generic patches can be inferred and applied in a pipeline of Automated Program Repair (APR).
With \toolname, we address an urgent challenge in the template-based APR community to separate implementation details from actual scientific contributions by providing an open, transparent and flexible repair pipeline on top of which all advancements in terms of efficiency, efficacy and usability can be measured and assessed rigorously.
Furthermore, because the underlying tools and concepts have already been accepted by a wide practitioner community, we expect \toolname's adoption by industry to be facilitated. Preliminary experiments with a prototype \toolname on the IntroClass and CodeFlaws benchmarks suggest that it already constitutes a solid baseline with comparable performance to some of the state of the art.
\end{abstract}

% \IEEEpeerreviewmaketitle

\section{Introduction}
\label{sec:intro}
In the race for achieving the old software engineering dream of automating program repair, approaches that leverage fix patterns currently have the lead (in terms of how many benchmark bugs can be fixed)~\cite{liu2020efficiency}. Unfortunately, despite the excitement of this momentum in the research community, practitioners expectations are not met and full integration in industrial settings remain anecdotical. Initial experimental attempts to large-scale application of automatic bug fixing suggest however that pattern-based patch generation fits the current practice of software engineering: \ding{182} At Facebook, Getafix~\cite{bader2019getafix} and SapFix~\cite{marginean2019sapfix} suggest fixes for in-house software by learning patterns using ``hierarchical clustering to many thousands of past code changes that human engineers made, looking at both the change itself and also the context around the code change''. \ding{183} In the Linux open source ecosystem, the Coccinelle~\cite{padioleau2008documenting} code transformation engine, which builds on pattern-like specifications written by developers, has been leveraged to automatically generate over 6\,000 patches~\cite{lawall2018coccinelle} that were accepted in the kernel code base. \ding{184} Besides repair, Ubisoft designed Clever~\cite{clever} to detect risky commits at commit-time using patterns of programming mistakes from the code history.

Recently, Liu et al.~\cite{liu2019tbar} proposed to revisit the performance of automated program repair (APR), carefully searching to identify the pattern databases that were available in the literature. Their experience report suggests that researchers do not actually share a common definition of what constitutes a repair pattern: levels of abstraction vary significantly and their immediate exploitation is often impossible as a transferable ingredient. Koyuncu et al.~\cite{koyuncu2020fixminer} and Ueda et al.
~\cite{ueda2020devreplay}, in two independent works, pointed out that fix patterns should  be made {\em tractable} (i.e., they should be a clearly identifiable artefact in the repair pipeline towards explaining the patch generation decisions) and {\em editable} (i.e., APR users should be enabled to intervene to correct these patterns manually to take into account project specificities). On top of these concerns, the full APR pipeline generally suffers from a lack of:
\begin{itemize}[leftmargin=*]
    \item {\bf Practicality}: A large body of the literature in APR present approaches that target well curated benchmarks with several constraints (e.g., test cases are readily available for the identified bugs) which may not be the case in practice. Although recent works~\cite{koyuncu2019ifixr} have started to investigate the use of bug reports, their pipelines remain heavily driven by test suites (e.g., for validation). 
    \item {\bf Flexiblity}: Regardless of the patch generation process (i.e., heuristic-based, constraint-based, or template-based following the taxonomy proposed in~\cite{le2019automated}), the available change transformations are generally limited to small mutations operators which are tightly embedded in the proposed algorithms. Seldom, an approach allows third party members in the community to readily edit and extend the list of possible code transformations.
    \item {\bf Transparency}: The repair approaches suggest patch candidates from  a search space. However, in most of the case, the origins of the patch candidates, i.e., how they are discovered, is missing. This intraceability remains a big obstacle for the transparency.
\end{itemize}

Overall, we note that on the road to automated program repair, the practitioner community is looking for techniques that can rapidly recommend patches that may be manually validated by developers. Indeed, these appear to be acceptable in various industrial settings so far. It may thus be worthwhile to drain some research effort into building an automatic patch generation system that is based on robust, agile and tractable techniques for inferring code transformation strategies.

{\bf This paper.} We present the core concept behind \toolname, a flexible, transparent and practical APR pipeline. We have initiated this \toolname and call for the community to commit on working on its building blocks for delivering a reliable tool support for practitioners in the context of program repair. \toolname is built on top of well-accepted software maintenance concepts in the Linux community, notably the concept of {\em generic patch} (more known as {\em semantic patch} in recall for the specification language: the Semantic Patch Language~\cite{brunel2009foundation}). {\bf We use {\em generic patch} specification as the tractable notation for fix patterns}.  The main contributions of our work are as follows:
\begin{itemize}
    \item We propose a critical review of template-based APR steps and suggest the design of a patch generation system around the concept of {\em generic patch} whose underlying definition and structure is borrowed from the Linux community toolbox.
    \item We initialize an open framework for program repair. The proposed pipeline, \toolname, transparently uses state of the art building blocks that can be customized.
    \item We evaluate the prototype pipeline, using available building blocks from the literature. Performance is measured with $C$ program repair benchmarks.
\end{itemize}
\section{Related Work}

\noindent
{\bf Program repair at a glance.}
%\tb{discuss the different kinds of program repair - taxonomy of Le Goues}
% \input{tables/APRTools}
Patch generation is one of the key tasks in software maintenance since it is time consuming and tedious. 
If this task is automated, the cost and time of developers for maintenance will inevitably be reduced in a dramatic manner. 
To address this issue, many automated techniques have been
proposed for program repair~\cite{le2019automated,monperrus2018automatic}. Ultimately, program repair is about traversing a search space of patch candidates that are generated by applying change operators to the buggy program code. 
Depending on how a technique conducts the search and constructs the patches, it can be considered as {\em heuristics-based}~\cite{jiang2018shaping,kim2013automatic,liu2019tbar,westley2009automatically}  or {\em constraint-based}~\cite{mechtaev2016angelix,nguyen2013semfix,xiong2017precise} following the taxonomy proposed by Le Goues {\em et al.}~\cite{le2019automated}. 
If such a technique further leverages learning mechanisms to infer transformation patterns, or to build patch models or to even predict patches, it is considered as {\em learning-aided}~\cite{gupta2017deepfix,long2017automatic,long2016automatic}.

In the last decade, most proposed techniques in the literature present repair pipelines where patch candidates are generated then validated against a program specification, generally a (weak) test suite. We refer to them as {\em generate-and-validate} test-suite based repair approaches and focus \toolname framework under this practical repair scheme. 
The genetic programming-based approach proposed by Weimer {\em et al.}~\cite{westley2009automatically}, as well as follow-up works, appeared only valid for hypothetical use cases.
Nevertheless, in the last couple of years, two independent reports have illustrated the use of literature techniques in actual development flows: in the open source community, the Repairnator project~\cite{urli2018design} has successfully demonstrated that automated repair engines can be reliable: open source maintainers accepted and merged patches which were suggested by an APR bot. At the premises of Facebook, the SapFix repair system has been reported to be part of the continuous integration pipeline~\cite{marginean2019sapfix} while Getafix was used there at large scale~\cite{scott2019getafix}.

Given fault localization information that pinpoints the code locations in the program that are the most likely to be buggy, test suite program repair approaches apply syntactic transformations to generate patches. 
Early techniques such as GenProg~\cite{claire2012genprog,westley2009automatically} relied on simple mutation operators to drive the genetic evolution of the code. More widespread today are approaches that build on fix patterns~\cite{kim2013automatic} (also referred to as fix templates~\cite{liu2018mining} or program transformation schemas~\cite{hua2018towards}) learned from existing patches.
Several APR systems~\cite{kim2013automatic,saha2017elixir,durieux2017dynamic,liu2018mining,hua2018towards,koyuncu2020fixminer,martinez2018ultra,liu2019avatar,liu2019you,liu2018mining2,liu2019avatar,liu2019tbar} implement this strategy by using diverse sets of fix patterns obtained either via manual generation or automatic mining of bug fix datasets. Unfortunately, whether they are generated on-the-fly (e.g.. with SimFix~\cite{jiang2018shaping} and CapGen~\cite{wen2018context}) or stored in a database, fix patterns remain an elusive concept.

\ding{43} In this work, our aim is to establish generic patches specified via the Semantic patch language as the formal notation for abstracting and defining fix patterns.

\vspace{1mm}
\noindent
{\bf Fix patterns inference for program repair.}
While the literature includes a large body of work on change patterns~\cite{fluri2008discovering,martinez2015mining,fluri2006classifying}, and more generally on change redundancies~\cite{pan2009toward,kim2009discovering, kim2006memories, molderez2017mining,yue2017characterization,nguyen2010recurring,liu2018closer}, very few approaches have actually leveraged again their ``discovered'' patterns to instantiate repair patches. 

Nevertheless, specific bug patterns have been mined to build fixing engines: Livshits and Zimmermann~\cite{livshits2005dynamine} discovered application-specific repair templates by using association rule mining on two Java projects while Hanam et al.~\cite{hanam2016discovering} have developed the BugAID technique for discovering most prevalent repair templates in JavaScript.

DevReplay~\cite{ueda2020devreplay} is a recent static analysis tool that suggests  source code changes based on a project's git history. The proposed changes can be edited by users without requiring knowledge about the AST.

FixMiner~\cite{koyuncu2020fixminer} is an automated approach to mining relevant and actionable fix patterns based on an iterative clustering strategy applied to atomic changes within patches. The goal of FixMiner is  to infer separate and reusable fix patterns that can be leveraged in other patch generation systems. This approach provides an appealing building block in the context of the \toolname framework. Unfortunately, its patterns are  also not immediately actionable; they must be manually integrated into a repair engine, which require a tedious an error-prone hard-coding of bug-fixing patterns. Additionally, FixMiner patterns do not contain any code token information: they have holes. The donor code should be searched before generating a concrete patch, which may lead to various non-sensical patches. 
FixMiner currently supports only Java and it does not provide any end-to-end traceability (i.e. we do not  know from where the pattern has been inferred).

\ding{43} We borrow some ideas from the FixMiner approach for computing patch similarity towards inferring patterns. In particular we find their rich AST edit script to be  appealing for building the prototype implementation of \toolname.

\vspace{1mm}
\noindent
{\bf Generic patches in the literature.}
There have been some work addressing the problem of considering a set of patches and attempting to find a ``generic patch'' that summarizes the change that is common across the patches. Chawathe et al. proposed a seminal method to detect changes to structured information based on an ordered tree and its updated version~\cite{chawathe1996change}. The goal was to derive a compact description of the changes with the notion of minimum
cost edit script which has been used in the recent ChangeDistiller and GumTree tools. Spdiff~\cite{andersen2010generic,andersen2012semantic} was then a promising approach that considered inferred change patterns from a set of patches. It was however found to scale poorly to a large number of patches, and to have constraints in producing ready-to-use patterns that can be used (e.g., by the Coccinelle matching and transformation engine~\cite{brunel2009foundation}).
Recently, Serrano et al.~\cite{serrano2020spinfer} proposed Spinfer as a tool-supported approach to ease large-scale changes across many source files in Linux by suggesting transformation rules to developers, inferred automatically from a collection of examples. 

\ding{43} Spinfer builds on the notation of ``generic patch'' (also referred to as ``semantic patch''), which Linux developers are already familiar with, thanks to the wide adoption of the Coccinelle~\cite{padioleau2008documenting} transformation engine and the associated Semantic Patch Language. We will rely on this building block for the inference of fix patterns in the prototype version of \toolname.

\begin{figure*}[!t]
	\centering
	\includegraphics[width=1\linewidth]{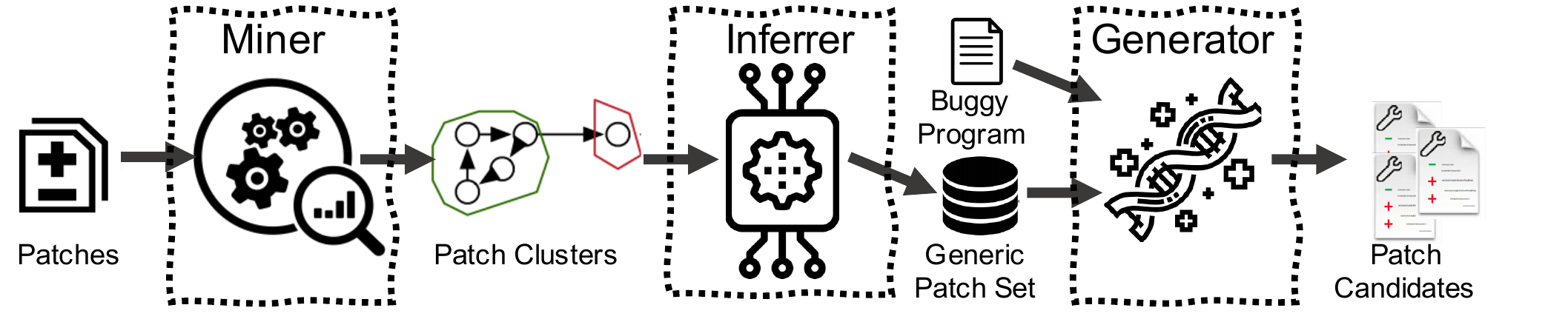}
	\caption{The \toolname pipeline.}
	\label{fig:pipeline}
\end{figure*}

\section{The \toolname~Framework}
\label{sec:approach}

\toolname builds on the momentum of template-based program repair, which has been shown successful in fixing a variety of bugs in APR benchmarks. To date these approaches are among the most effective (in terms of the number of benchmark bugs that are fixed) repair tools in the literature. Relevant approaches in the literature (e.g., TBar~\cite{liu2019tbar}, AVATAR~\cite{liu2019avatar}, CapGen~\cite{wen2018context}, SimFiX~\cite{jiang2018shaping}) are often provided in a monolithic tooling which prevents extension, adaptation and even application on real-world code bases beyond those targeted by initial experimental validations. 

\note{In this work, we propose to initiate a community-wide effort to build a flexible, transparent and practical framework for template-based program repair to (1) enable better assessment of research advancements, and (2) facilitate adoption of APR by software maintainers. 
}

\toolname is  carefully designed to ensuring that its users have control over important steps of the patch generation process. In particular, we consider the following critical questions:

\noindent
\ding{182} {\bf Where should we mine repair transformations?} Template-based program repair systems, whether they leverage specifically pre-defined mutation operators, infer code transformations on-the-fly or rely on offline-inferred fix patterns, they generally build on data of existing code bases (preferably with a large history of code changes). If the source of mining is not appropriate (e.g., limited recurrent changes or changes associated to domain-specific bugs), the mined patterns may be irrelevant for the program that is targeted for repair.

\noindent
\ding{183} {\bf How are fix patterns inferred?} A challenge that has been highlighted in two recent independent studies~\cite{koyuncu2020fixminer, ueda2020devreplay} is that fix patterns discussed in the APR community are largely intractable artefacts. If the underlying fix patterns cannot be manipulated (i.e., checked and edited) by practitioners, the adoption of the integrating APR tool will be largely hindered.

\noindent
\ding{184} {\bf How are patches generated?} Besides fault localization information which generally drives the selection of fix patterns, the application of code transformations generally follows various ad-hoc recipes and involve empirical design choices for fix pattern matching and donor code search. If these activities cannot be ensured to be deterministic, industry adoption of APR cannot be ensured. 

\subsection{Execution steps of \toolname}
We propose to build an APR pipeline that addresses the issues raised in the aforementioned questions. 
Figure~\ref{fig:pipeline} illustrates an overview of the \toolname.

The pipeline takes the code repositories that the maintainer judge to be relevant for learning code transformations as its input. This set of code repositories can be constituted by the single source code repository associated to the program under repair. 
Then, each of the questions formulated above are addressed by a major component involved in a specific step of the \toolname pipeline :
\begin{itemize}
    \item \miner analyzes the structural similarity between input repository patches and yields clusters that can be tuned by \toolname users to take into account the recurrence level of code transformations that will be supported by the patch generation. 
    \item \inferrer then abstracts fix patterns from each retained cluster and specifies it in a format that can be inspected (for relevance) and edited (to account for specific maintenance style requirements).
    \item \generator finally builds the concrete patches for the given buggy programs, after attempting to match fix patterns to the appropriate code locations (i.e., the likely buggy code locations).
\end{itemize}

Instead of re-inventing new algorithms and prototyping tools that would require extensive vetting before adoption, we propose to bootstrap the \toolname pipeline by relying on tried-and-true technologies that software maintenance are already familiar with. 
Concretely, we have identified a code transformation tool that is part of the Linux kernel developer toolbox since 2008 and which is now increasingly used to automate large-scale changes in kernel code. 
This tool, Coccinelle~\cite{padioleau2008documenting} builds on a concept of {\em semantic patch} that allows developers to write transformation rules using a diff-like syntax. {\bf In this work, we will use instead the term ``{\em generic patch}'' to refer to the specification of transformation rules that can be given as input to Coccinelle.}
A generic patch is thus an abstraction that uses metavariables to represent common but unspecified subterms (e.g., any variable) and notation for reasoning about control flow paths.

Given the standing of Linux development practices in the software development community, the adoption of a tool such as Coccinelle, and its underlying concepts, is a strong signal that it fits with industry standards. We therefore propose to build the \toolname pipeline on top of the Coccinelle engine. 

\note{A fix pattern in \toolname is a {\tt generic patch} that is specified using the specification language of Coccinelle, which is now integrated to the Linux development toolbox. }

%  Then, in Step 1, we mine code changes
% that are similar in terms of context and operations. Our technique, extends  Fixminer~\cite{koyuncu2020fixminer}, to discover relevant fix patterns from similar changes based on different tree representations encoding contexts, change operations. Fixminer forms cluster of patches that are sharing a common tree representations. In Step 2, we leverage Spinfer~\cite{serrano2020spinfer} to identity transformation rules that are inferred automatically
% from the clusters of the patches formed in Step 1. Lastly in Step 3, we use Coccinelle ~\cite{padioleau2008documenting} to generate the concrete patches from the templates. 

% In the following sections, we present the details of each step.

\subsection{Overview of the SmPL Language }
\label{sec:smpl}
The Coccinelle tool is an example of public research effort that gained traction in industrial settings, thanks to support from the open source community. It was initially designed to document and automate {\em collateral evolutions}~\cite{padioleau2008documenting} in the Linux kernel source code, but is now used in a variety of other code bases as an base engine for performing control-flow-based program searches and transformations in C code~\cite{lawall2018coccinelle}. Coccinelle integrates a static analysis that
is specified using control-flow sensitive concrete syntax matching rules. Search (i.e., identifying code fragments that match a pattern) and transformations (i.e., generating patches following the fix pattern) are specified via the Semantic Patch language (SmPL) language, and executed by a dedicated transformation engine.  Although the Linux community refers to the SmPL specifications as "semantic patches", we will refer to them in \toolname as "generic patches" to reflect the idea that they are abstract patterns that must be "concretized" into generated patches.

Although SmPL specifications can contain OCaml or Python code, allowing to perform arbitrary computations, in this work we focus on its pattern matching and code transformation capabilities. 
Listing~\ref{lst:smpl1} provides an example of generic patch (as an SmPL specification). The patch goal is :
\begin{itemize}
    \item (1) to identify all code locations where there is an attempt to access a field of struct whose pointer has not been safely checked beforehand in the control flow. Indeed if the pointer ($param$) is NULL, the dereference would lead to a segmentation fault (and a crash in the case of an operating system code).
    \item (2) to produce a corrective patch (i.e., adding check and early return statement) at all places where such an unsafe derefence can take place. 
\end{itemize}

The generic patch is constituted of a single rule named ``unsafe\_dereference'' which defines five metavariables (lines~2-4): $T$ (type) which represents any data type; $p$, which represents an arbitrary position in the source program; $fn$  (function name), $param$ (parameter name) and $fld$ (data structure field name), which represent arbitrary identifiers. {\em Metavariables} are bounded by matching the code pattern against the source code. For example, the pattern
fragment on line 6 ({\tt fn(..., T *param, ...)}) will cause fn to be bounded to the name of a function in its definition, and cause $param$ to be bounded to any pointer parameter name. The notation $@p$ binds the position metavariable $p$ to information about the position of the match of the preceding token. Once bounded, {\bf a metavariable must maintain the same value within the
current control-flow path}; thus, for example, the occurrences of param on lines~6-13 must all match the same expression. The fix pattern (lines~6-15) therefore consists of essentially C code, mixed with a few operators to raise the level of abstraction so that a single generic patch can apply to many code sites.

\definecolor{codegreen}{rgb}{0,0.5,0}
\definecolor{codegray}{rgb}{0.5,0.5,0.5}
\definecolor{codepurple}{rgb}{0.58,0,0.82}
\definecolor{backcolour}{rgb}{0.95,0.95,0.92}

\lstdefinestyle{cocci}{
    backgroundcolor=\color{backcolour},   
    commentstyle=\color{codegreen},
    keywordstyle=\color{magenta},
    numberstyle=\tiny\color{codegray},
    stringstyle=\color{codepurple},
    basicstyle=\ttfamily\small,
    breakatwhitespace=false,         
    breaklines=true,                 
    captionpos=b,                    
    keepspaces=true,                 
    numbers=left,                    
    numbersep=5pt,                  
    showspaces=false,                
    showstringspaces=false,
    showtabs=false,                  
    tabsize=2,
    keywordstyle=\color{black}\bfseries, 
    morekeywords={when,type, position,expression, identifier, parameter,any,exists}
}

\lstset{style=cocci}

\begin{lstlisting}[escapechar=^, caption=Example of generic patch,label={lst:smpl1},stringstyle=\scriptsize\ttfamily,basicstyle=\scriptsize\ttfamily,aboveskip=1pt,belowskip=1pt]
@unsafe_dereference exists^@^
type T;
position p;
identifier fn, param, fld;
@@
fn(.., T *param, ...){
... when != param = new_val
    when != param == NULL
    when != param != NULL
    when != IS_ERR(param)
^\textcolor{codegreen}{+~~~if (param == NULL)}^
^\textcolor{codegreen}{+~~~~~return}^
    param->fld@p
... when any
}
\end{lstlisting}

\subsubsection{Sequences abstraction}
The main abstraction operator provided by SmPL is `...', representing a sequence of
terms. In line~6, `...' represents the remaining parameters of a function that appear before and after a given parameter is matched in the parameter list; in line 7, `...' represents the sequence of statements reachable from the begin of the definition of a function along any control-flow path. By default\footnote{This default behavior can also be explicitly stated using the ``{\bf \tt forall}'' annotation}, such a sequence is quantified over
all paths (e.g., over all branches of a conditional block), but the annotation ``{\tt exists}'', next to the rule name, indicates that for this ``unsafe\_dereference'' rule, the matching should be done even for one path. It is also possible to restrict the kinds of sequences that `...' can match using the keyword {\bf \tt when}. Lines 9-12 use {\bf \tt when} to indicate that there should be
no reassignment of param nor any check on the validity of the param pointer value before reaching
the dereference that consists in accessing a field fld in the corresponding data structure.

A SmPL rule only applies when it matches the code completely. Consider the example of buggy code in Listing~\ref{lst:code}. The rule {\tt unsafe\_dereference} matches the parameter of type struct person * on line~1 and the dereference on line~6 as it exists a control-flow path where the validity of {\em pers} is not checked. The metavariable {\em fn} (cf. Listing~\ref{lst:smpl1} )is bound to the identifier {\em get\_age}, and the metavariable {\em param} is bound to {\em pers}. The metavariable {\em p} is
bound to various information about the position of the dereference, such as the file name, line number, character number (on the line).

\begin{lstlisting}[escapechar=^, language=c, caption=Example of buggy code with an unsafe dereference,label={lst:code},stringstyle=\scriptsize\ttfamily,basicstyle=\scriptsize\ttfamily,aboveskip=1pt,belowskip=1pt]
int get_age(int alive, struct person *pers, char *context){
int age=0;
if (alive == 1 && pers !=NULL)
    age=pers->age_death - pers->age;
else
^\textcolor{red}{~~~ age = pers->age;}^
return age;
}
\end{lstlisting}

\subsubsection{Disjunctions and Nests}
Besides the `...', SmPL provides \textbf{disjunctions}, $(pat_1\;|\;...\;|\;pat_n)$, and
\textbf{nests}, $ <...\;pat\;...>$. A nest $< ...\;pat\; ... >$ in SmPL matches a sequence of terms, like `...’. However, additionally, it can match zero or more occurrences of \textit{pat} within the matched sequence. Another form of
nest exists for matching one or more occurrences of pat. By analogy to the + operator of regular
expressions, this form is denoted $<+...\;pat\; ...+>$.

The examples discussed above illustrate the abstraction power that generic patches provide in the activity of propagating fixes. In the next section we present the steps for:
\begin{itemize}[leftmargin=*]
    \item {\em \bf Regrouping patches} into clusters where bug fix code transformations are made with similar patterns.
    \item {\em \bf Automatically generating generic patches} from clusters of patches in order to populate the repair template databases.
    \item {\em \bf Performing patch generation} in practice given an identified buggy code location (even at a coarse granularity)
\end{itemize}

\subsection{Patch clustering}
\label{subsec.patching}

The goal of the \miner is to perform patch clustering, i.e.,  to group together the code changes that are representing a repeating code context and change operations. 
In order to convey the full syntactic and semantic meaning of the code change and to discover cluster of patches that are sharing a common representation, we leverage the rich AST edit script representation proposed by Koyuncu et al.~\cite{koyuncu2020fixminer}. 

% we resort to \fixminer.\ak{TODO} \fixminer is an automated mining approach to discover cluster of patches that are affecting the same code context and the same repair actions. It uses a specialized tree structure of the edit scripts, refered as Rich Edit Script, which captures the AST-level context of the code change. 

\begin{Grammar}[!h]
    \centering
    \scriptsize
    \vspace{2mm}
    \begin{grammar}
<richASTEditScript> $\rightarrow$ <node>+ 

<node> $\rightarrow$ `-\,-\,-'* <move> | `-\,-\,-'* <delete> | `-\,-\,-'* <insert> | `-\,-\,-'* <update> 

<move> $\rightarrow$ `MOV' <astNodeType> @@ <tokens> `@TO@' <astNodeType> `@@' <tokens> `@AT@'

<delete> $\rightarrow$ `DEL' <astNodeType> `@@' <tokens> `@AT@'

<insert> $\rightarrow$ `INS' <astNodeType> `@@' <tokens> `@TO@' <astNodeType> `@@' <tokens> `@AT@'

<update> $\rightarrow$ `UPD' <astNodeType> `@@' <tokens> `@TO@' <tokens> `@AT@'

\end{grammar}
    \caption{Notation of rich AST edit script}\label{gra:rich}
    \vspace{0.2cm}
\end{Grammar}
A rich AST edit script, whose grammar is illustrated in Grammar~\ref{gra:rich}, encodes the information about the AST node types in a change diff tree, the repair actions performed, the raw tokens involved as well as the parent-child relation among the nodes. 
We consider only code context and change operation presentation (cf. Figure~\ref{fig:fixminerPattern} ) to detect similar changes and group them into clusters of similar code changes. The objective of this step is to ensure that we can reduce the noise in pattern inference, regrouping together patches that perform similar changes actions, and potentially filtering out cases where the redundancies of changes are limited.

% Given a patch, we start by computing the set of edit actions (edit script) using GumTree, where the set contains an edit action for each contiguous group of code lines (hunks) that are changed by a patch. 

% \fixminer uses a folding mining strategy to leverage different information encoded in rich edit script, focusing on a single information type in each fold. It starts by mining the changes affecting the same code context, then, changes using the same repair actions are regrouped and finally, changes affecting the same tokens set are mined. In each fold, it compares the code snippets and detects changes having an identical presentation and groups those identical changes into clusters.
% Even though \fixminer is handy for discovering diverse sets of patch cluster, the lack of support for $C$ programs and the limited traceability information and tunability of parameters prevents a direct usage of \fixminer in \toolname pipeline. 
% In this work we extend \fixminer in order to overcome the limitations aforementioned. Concretely, we extend the Rich Edit Script to srcML~\cite{maletic2002source} parser in order to add $C$ language support, which was previously based on JDT parser. 

\begin{figure}[!ht]
    \centering
    \vspace{2mm}
    % \lstinputlisting{patternListings/black/expr_stmt_8_12.cocci4-prev}
\begin{lstlisting}[stringstyle=\scriptsize\ttfamily,basicstyle=\scriptsize\ttfamily,aboveskip=1pt,belowskip=1pt]
UPD expr_stmt 
--- UPD expr 
------ UPD call 
--------- UPD name 
--------- UPD argument_list 
------------ DEL argument 
--------------- DEL expr 
------------------ DEL literal:string 
\end{lstlisting}
    \caption{An example code context and change operation presentation.}
    \label{fig:fixminerPattern}
\end{figure}
% The three available mining iterations (i.e, code context, repair actions and tokens) of \fixminer is simplified into a single iteration where we consider only the changes that are sharing the same code context and repair actions, illustrated with an example in  Figure~\ref{fig:fixminerPattern}. Furthermore, we add configuration parameters to set the mining preferences (i.e, parameters to limit the number of changed lines in a patch, the number of hunks in the patch etc. ) to provide a tunable interface to users. And finally, we make some minor changes to improve the traceability (i.e, the origins of the code snippets that are forming the patch cluster).

\subsection{Generic Patch Inference}
\label{subsec.inference}
The goal of the \inferrer is to derive generic patches from the clusters of similar concrete patches that have been mined in the previous step. We build on a recent work by Serrano et al.~\cite{serrano2020spinfer} which showed that it is possible to automatically generate SmPL transformation rules by learning from examples. The approach  considers both similarity among code fragments and among control flows associated to the changes to identify
change patterns and specify transformation rules.

In practice, the idea is to abstract over common changes across the examples, incrementally extending a pattern until obtaining a rule that describes the complete change, respecting both control-flow and data-flow relationships between the fragments of the code. To that end, each patch in a cluster is used to reconstitute the before- and after-change files, then the \inferrer identifies sets of common removed or added terms across the examples, and further generalizes these terms in each set into a pattern that matches all of the terms in the set, and finally integrates these patterns into transformation rules that respect both control-flow and data constraints exhibited by the examples. 

Figure~\ref{fig:exampleSpinfer} illustrates an example of inferred generic pattern. It encodes the information of how to transform code, the locations of where the pattern is inferred, as well as statistics on the recall (i.e., the percentage of expected changes in the examples that are obtained by applying the inferred generic patch), precision (i.e., the percentage of changes obtained by applying the inferred generic patch that are identical to the expected changes in the examples). We benefit in \toolname from the  SmPL-specified generic patch notation which is close to C: it makes the patterns understandable to the user and even allows the user to improve the script or adapt it to other uses, hence contributing to {\bf transparent} schema of program repair where patterns are tractable.
\begin{figure}[!ht]
    \centering
    % \lstinputlisting{patternListings/black/expr_stmt_8_12.cocci4-prev}
\begin{lstlisting}[stringstyle=\scriptsize\ttfamily,basicstyle=\scriptsize\ttfamily,aboveskip=1pt,belowskip=1pt]
@@
identifier I0;
@@
- char I0; 
+ int I0; 
// Infered from: (MultiMarkdown-6/{prevFiles/prev_626381_73d644_Sources#libMultiMarkdown#aho-corasick.c,revFiles/626381_73d644_Sources#libMultiMarkdown#aho-corasick.c}: ac_trie_search)
// Recall: 0.11, Precision: 1.00, Matching recall: 0.50
\end{lstlisting}
    \caption{An example of generic patch.}
    \label{fig:exampleSpinfer}
\end{figure}

\vspace{-4pt}
\subsection{Code Transformation with Generic Patches}
\label{subsec.semantic}
Given that we leverage the SmPL language to specify generic patches, code transformation is provided for free by the Coccinelle search and transformation engine. The engine takes as input a generic patch and a source code file that it parses. Then it performs a control-flow matching to identify code locations whose shape fit with the structure of the code structure targeted by the generic patch. Taking the metavariables values bounded at each matched code location, it then generates the necessary concrete patches. This engine thus provides two essential advantages over existing generate-and-validate repair pipelines: 
\begin{itemize}[leftmargin=*]
    \item (1) there is no need for a fine-grained bug localization engine. A coarse-grained localizer that points to buggy files can be leveraged. Thus, even IR-based bug localization tools which produce results at file level without requiring test cases can be relevant (as advocated by recent work~\cite{koyuncu2019ifixr}). 
    \item (2) the search for donor code is facilitated by the use of metavariables in the generic patch, allowing for the transformation engine to infer and track tokens across the control-flow, and thus maximizing the chances of producing sensical patches (i.e., patches that at least make the program compile).
\end{itemize}

\section{Study Design}
\label{sec:studyDesign}

We now overview some details of our experimental validation. The objective in this study is not to build a novel state of the art repair tool, but to rather offer a new perspective into a framework for template-based program repair with at its core the concept of generic patch for specifying fix patterns (aka fix templates).
Before presenting the results, we discuss the dataset of code repositories that we build for mining similar patches and further inferring generic patches.  Then, we present the benchmarks used to assess the overall performance of \toolname as a our prototype implementation.
Finally we overview the implementation choices made for both the patch clustering and pattern inference steps.

\subsection{Subjects}

\toolname provides a flexible interface to its user, 
who can decide either to use a specific code repository to mine the fix patterns, or to use the  pre-constructed fix pattern database shipped with the framework. 
To build this database we needed to first collect a large set of the code repositories with a long history of code changes. 
%It can either use a code repository specified by the user to mine fix templates or the predefined fix template dataset shipped with the pipeline. 

The subjects that are included in our dataset are collected: i) by manual identification of popular $C$ repositories from gihub, gitlab, savannah with a large code history and ii) systematically by leveraging the build activities in Travis CI. For the latter we refer to the data of Durieux et al~\cite{durieux2019analysis}, which contains all the Travis CI jobs executed between 30 September 2018 and 22 January 2019 by 272\,917 projects. From their dataset we identified 2\,858 $C$ repositories. We further curates this dataset based on the repository properties from github (i.e., commit count, watchers count, forks count etc.). Eventually we select the repositories i) that are not forks, ii) having at least 200 commits, iii) at least 10 watchers iv) and at least 10 forks. Our dataset eventually includes 351 repositories. Table~\ref{tab:dataset} lists some of the major ones.
\begin{table}[!ht]
	\centering
	\caption{Some of selected repositories used in our study.}
	\resizebox{.9\linewidth}{!}
	{
	\begin{threeparttable}
		\begin{tabular}{l|r|rr}
\toprule
\multicolumn{4}{c}{Systematic identification}\\
\midrule
repository &  commitCount &  watchers &  forks \\
\midrule
                           xqemu/xqemu &        68836 &       462 &     49 \\
                               git/git &        59910 &     33398 &  19513 \\
                     greenplum-db/gpdb &        56052 &      4073 &   1171 \\
                 MonetDB/MonetDBLite-C &        54946 &        26 &     13 \\
                        panda-re/panda &        54869 &      1640 &    393 \\
        %   FreeRADIUS/freeradius-server &        35280 &      1257 &    792 \\
        %             isc-projects/bind9 &        32164 &       157 &     58 \\
        %              kamailio/kamailio &        31474 &      1179 &    585 \\
        %                  open-mpi/ompi &        30888 &      1054 &    549 \\
        %                 openpmix/prrte &        30679 &        12 &     28 \\
        %               openssl/openssl &        26592 &     13439 &   5964 \\
                       
\midrule                    
\multicolumn{4}{c}{Manual identification}\\
\midrule
repository &  commitCount &  \multicolumn{2}{c}{source} \\
\midrule
linux & 949406 &  \multicolumn{2}{c}{git.kernel.org} \\
freebsd & 271000 &  \multicolumn{2}{c}{github.com}\\
FFmpeg & 98901 & \multicolumn{2}{c}{github.com}   \\
cmake  &49611 & \multicolumn{2}{c}{gitlab.kitware} \\
gtk &65679 & \multicolumn{2}{c}{gitlab.gnome} \\
\midrule

\end{tabular}

	\end{threeparttable}
	}
	\label{tab:dataset}
\end{table}
\vspace{-4pt}
\subsection{Assessment Benchmarks}
 We selected the Introclass~\cite{le2015manybugs} and Codeflaws~\cite{tan2017codeflaws} datasets to  empirically assess \toolname.

The Introclass dataset is a benchmark of small $C$ programs collected from classroom assignments of students. It includes 998 defects, 778 of them being associated with an instructor constructed black-box test suite and 845 being associated with a white-box test suite created using KLEE~\cite{cadar2008klee} (a symbolic execution tool that automatically generates tests). 

Codeflaws benchmark consists of 3\,902 defects collected from C programs developed during programming contests. The benchmark is associated with two sets of test-suites: i) a test suite given to repair tools for generating repair ii) a held-out test suite for validating the correctness of patches. 

Table~\ref{tab:benchmarks} lists some statistics about the benchmarks.

\begin{table}[!ht]
	\centering
	\caption{Basic statistics of Benchmarks.}
	\resizebox{\linewidth}{!}
	{
	\begin{threeparttable}
		
\begin{tabular}{l|rrrr}
\toprule
        Benchmark    &  \# of Defects & Size of Test Suite I & Size of Test Suite II & LOCs \\
\midrule
        Codeflaws    & 3902    & 2-8 & 5-350 & 1-322 \\
        Introclass    & 998    & 6-9 & 6-10 & 13-24 \\

\bottomrule
\end{tabular}
	\end{threeparttable}
	}
	\label{tab:benchmarks}
\end{table}

\subsection{Implementation Choices}
\toolname aims for flexibility and extensibility such that practitioners may tune parameters and adapt the framework to their requirements. 
We recall that we have made the following parameter choices in the \toolname:
\begin{itemize}
	\item Repository selection is made based on the C programs with a large commit history and which are actively used.
	\item Change size in a patch is limited to have at most 50 changed lines.
	\item Patch spread is limited such that each patch contains at most 3 hunks.
	\item Timeout for generic patch inference is set to 900 seconds for each patch cluster. 
\end{itemize}
% We extracted 350676 code snippets, having at most 50 changed  lines  in  a  patch, and at most 3 hunks.

% % \input{experiment}
\section{Assessment}
\label{sec:exp}

We assess the prototype framework of \toolname via performing experiments that answer the following research questions.  

\subsection{Research Questions}

%The assessment experiments are performed with the objective of investigating the usefulness of \toolname.
%\jk{In Intro, we claim that our pipeline is practical, flexible and transparent. Should we assess these properties? } 
%\jk{We can also target other criteria. For instance, in Section III, we have 3 questions.  }
%\jk{Or we simply say that we want to assess the performance of each of the 3 main components of \toolname: Miner, Inferrer, and Generator. }
%To that end, we focus on the following research questions (RQs):
\begin{enumerate}

    \item[{\bf RQ-1}:] {\bf\em To what extent can the application of \miner and \inferrer produce generic patches from the collected code repositories?} 
    %To that end, we first assess the relevance of the patch clusters yielded by the Miner component of \toolname. Then, we assess the \jk{quality?} if the fix patterns yielded by Inferrer.

	\item[{\bf RQ-2}:] {\bf\em Where did \toolname find relevant redundant changes to mine the generic patches?} 
	
	\item[{\bf RQ-3}:] {\bf\em What is the repairability performance of \toolname? }
	
% 	\item[{\bf RQ5:}] {\bf\em : Are patterns inferred by \toolname compatible with known fix patterns? }
		
	\item[{\bf RQ-4}:] {\bf\em What is the efficiency performance of \toolname? }

\end{enumerate}

\section{Results}
\subsection{Generic Patch Inference Capability}
We first assess the relevance of the patch clusters yielded by the \textsc{Miner} component of \toolname. 
Then, we look at the generic patches yielded by the  \textsc{Inferrer} implementation. 

\noindent
\textbf{\textsc{Miner} Assessment.} Performance of \miner is evaluated through the clusters that it yields. The objective is to estimate whether it can find enough cases of recurrent changes within patches collected from project repositories to form clusters. A given patch cluster will contain all the patches having the similar code change hunks.  
%Since a given code hunk may represent a repeating context the clustering
%process groups together only tree pairs \jk{of what?} that are identical among themselves.
Table~\ref{tab:clusterStats}  overviews the statistics of clusters yielded by \miner by taking as input the dataset of repositories presented in Section~\ref{sec:studyDesign}. The implementation choices presented in Section~\ref{sec:studyDesign} are also followed. 
Overall, \num{350\,676} code hunks have been extracted. 
Among these hunks, we noticed that \num{110\,949} ($\sim$32\%) are unique code hunk, and thus, they cannot be part of a cluster.
For the remaining \num{239\,727} code hunks ($\sim$68\%), there exists at least one other code hunk, among the \num{350676} code hunks, which is identical. They can thus form clusters of more than one patch. 
Overall, among these \num{239\,727} code hunks, we identified \num{31\,310} patch clusters (i.e, the code hunks of each patch of a cluster are identical).

%This means that \num{110949} (\num{350676}-\num{239727}, $\sim$32\%) code hunks are unique. 
%qualified as a member of a cluster \jk{I do not understand}. Recall that, a cluster is formed when it has at least two members sharing a common representation.
\begin{table}[!ht]
	\centering
	\scriptsize
	\caption{Statistics on Patch Clusters  .}
	\resizebox{.8\linewidth}{!}
	{
	\begin{threeparttable}
		\begin{tabular}{c|c|c|c}
		\toprule
		     % & \multicolumn{2}{c|}{\# Change Hunks} &\makecell{\# Clusters} \\ \cline{1-3}
		     %& \multicolumn{2}{c|}{\# Change Hunks} &\makecell{\# Clusters} \\ \cline{1-3}
			Total \# of     & \# unique  & \# hunks which can form a   & \# clusters \\
			 hunks   &     hunks      &  cluster of at least 2 patches &  \\
			  \hline
       % \noalign{\smallskip}\hline\noalign{\smallskip}
           \num{350676} & \num{110949} &  \num{239727} & \num{31310} \\

		%\noalign{\smallskip}
		\bottomrule

		\end{tabular}
	\end{threeparttable}
	}
	\label{tab:clusterStats}
	
\end{table}

\begin{figure}[h]
	\centering
	\includegraphics[width=0.9\linewidth]{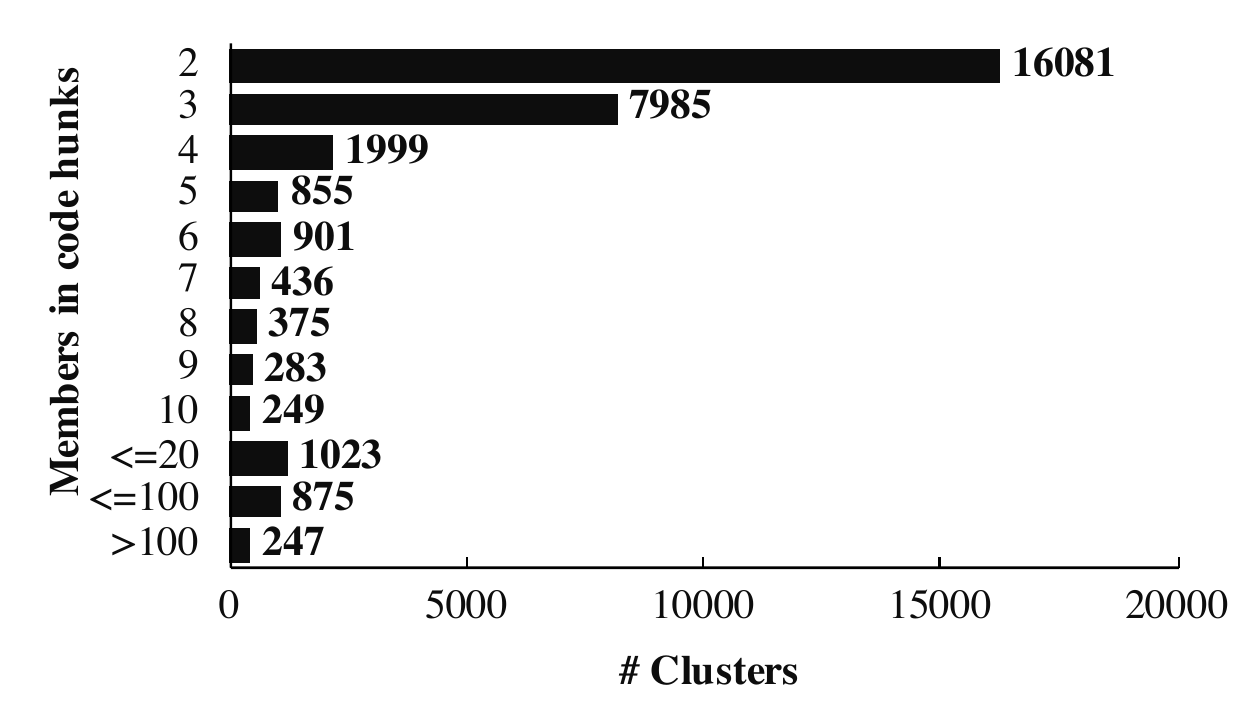}
	\caption{Distribution of the patch cluster sizes.}
	\label{fig:clusterMembers}
\end{figure}

Figure~\ref{fig:clusterMembers} shows the size distribution of the patch clusters. 
A majority of clusters, i.e.,  \num{16\,081} ($\sim$51\%), are formed by two recurrent code change hunks only. 
Conversely, \num{2\,394} (249+1\,023+875+247, $\sim$7.6\%) clusters contain at least 10 recurrent code change hunks. 

We further investigate how the cluster elements are spread across patches. 
To that end, we follow the categorization proposed by Koyuncu et al.~\cite{koyuncu2020fixminer}.
%To measure the spread among the patch we categorized the clusters formed into the following families:
\begin{itemize}[leftmargin=*]
	\item A {\em vertical cluster} is a cluster whose code change hunk is recurrent within a single patch. Such clusters are generally formed when we have patches that developers commit to perform a single type of change (e.g., change $kmalloc$ call to $kzalloc$ calls) across several code locations.
	\item An {\em horizontal cluster} is a cluster whose code change hunk is recurrent across several patches. Such clusters are formed when a code change (e.g., add a missing NULL check) is implemented by different developers for different code locations. 
\end{itemize}

Table~\ref{tab:clusterPatternSpread}  overviews the statistics of clusters yielded. 
Most of the clusters (24\;230) are horizontal clusters. This suggests that the same code changes are often spread among different patches, any or all of which may be used to infer the common generic patch. The vertical clusters can also be useful for inferring generic patches: they represent large patches making the same changes at once at several locations (e.g., collateral evolutions in Linux are applied through vertical patches~\cite{padioleau2008documenting}).

\begin{table}[h]
	\centering
	\caption{Statistics on Patch Clusters Spread}
	\resizebox{.8\linewidth}{!}
	{
	\begin{threeparttable}
		\begin{tabular}{lrrr|rrr}
		\toprule
				& \multicolumn{3}{c}{Vertical } & \multicolumn{3}{c}{Horizontal}  \\\cline{2-7}
			   & \makecell{\# Clusters} &\makecell{\# Patch} &\makecell{\# Hunk} & \makecell{\# Clusters} &\makecell{\# Patch} &\makecell{\# Hunk} \\
			  \hline
       % \noalign{\smallskip}\hline\noalign{\smallskip}
          & 3178&  3178 & 7565 & 24230 & 75691  & 75691\\
        \bottomrule
        % \hline
        % Shape-based  & 120 &  223 & 278 \\
        % Action-driven & 10 &  20 & 25  \\
        % Token-specific & 14 &  22  & 32 \\

		% %\noalign{\smallskip}
		% \hline
  %       Shape-based  & 47&  56 & 127 \\
  %       Action-driven & 6&  14 & 22  \\
  %       Token-specific &3 &  5  & 6 \\
  %       \hline
  %       Shape-based  & 9 &  17 & 21 \\
  %       Action-driven & 3 &  9 & 9  \\
  %       Token-specific & 3 &  5  & 6 \\

		\end{tabular}
		\begin{tablenotes}

    \item[*]
 A generic patch can simultaneously be vertical (when it is associated to several changes in hunks of the same patch) and horizontal (when it appears as well within other patches).
 \end{tablenotes}
	\end{threeparttable}
	}
	\label{tab:clusterPatternSpread}
	
\end{table}

\note{{\bf RQ-1.1}: \miner is practical. It is able to identify patch clusters (i.e., recurrent patch sets) of various sizes.}

\noindent
\textbf{\textsc{Inferrer} Assessment.}
We assess the ability of \inferrer to analyses changes within patch clusters and derive a generic patch (i.e., abstract the relevant fix pattern and specify it with the SmPL notation). Our implementation uses \spinfer as a backend for matching control-flow similarities. Preliminary experiments revealed that the approach is sensitive to the noise among patches. We expect our \miner step to have provided homogeneous patch clusters.
%{\bf Objective.} Initially, we propose to assess the fix pattern inference in terms of statistics on the inference process to provide a basis to evaluate whether they represent recurrent changes in the projects history.

%To infer fix patterns we used the patch clusters yielded which are already containing a set code hunks having a common representation.

Table~\ref{tab:inferredPatternStats}  overviews some statistics on the inferred generic patches.
From the \num{31310} patch clusters obtained with \miner, \inferrer was able to successfully yield a generic patch for 
\num{20467} ($\sim$65\%) clusters. 
The remaining clusters ($\sim$35\%) do not lead to any generic patch either because of the timeout value of 900 seconds set for analysing each patch cluster, or because they do not exhibit the necessary data or control-flow dependencies to satisfy any inference. 
Note that the initial generic patch inferred from a given cluster can contain several rules. We consider each transformation rule as a generic patch on its own. Eventually, we are left with \num{68\,368} atomic generic patches (i.e., generic patches with a single transformation rule).  

Note that in the middle column of Table~\ref{tab:inferredPatternStats}, we also report the number of code hunks that have been  used to infer the generic patches. Overall, \num{125483} ($\sim$52\%) out of \num{239727} code hunks contributed to a generic patch.

\begin{table}[h]
	\centering
	\caption{Inferred Generic Patch statistics.}
	\resizebox{.8\linewidth}{!}
	{
	\begin{threeparttable}
		\begin{tabular}{lr|r|r}
		\toprule
			   & \makecell{\# Patch Clusters} &\makecell{\# Code hunks} &\makecell{\# "Atomic" Generic Patches}  \\
			  \hline
       % \noalign{\smallskip}\hline\noalign{\smallskip}
          & 20\;467 &  125\;483 & 68\;368 \\

		%\noalign{\smallskip}
		\bottomrule

		\end{tabular}
	\end{threeparttable}
	}
	\label{tab:inferredPatternStats}
\end{table}

% Most of the inferred patterns contain multiple rules in the semantic patch. This is mainly due to the following observations:
% \begin{itemize}
% 	\item The patch cluster is a horizontal cluster, which may contains various implementation of the same changes that are using different variable names, expressions etc.. thus cannot be abstracted in a single rule. 
% 	\item \spinfer works with files before and after some change, on the other hand patch cluster are formed at code hunks several. Thus in the inference process \spinfer may infer a recurrent code hunk, which was not intended by the patch cluster.

% \end{itemize}
% In order to effectively use the patterns and avoid any duplicates we curate the semantic patterns to contain a single rule. Overall, our dataset contains 68\;368 unique semantic patterns. 

Table~\ref{tab:patterns} lists five of the most frequently observed generic patches in our dataset. 
We further manually investigate these generic patches by checking the corresponding commits in the repositories in order to understand the nature of the changes described by the developers. 
% \begin{figure}[!ht]
%     \centering
%     \lstinputlisting[stringstyle=\footnotesize\ttfamily,columns=fullflexible,basicstyle=\footnotesize\ttfamily]{patternListings/if_stmt_8_8.cocci1Example}
%     \caption{An example patch from generic patchg if\_stmt\_8\_8.}
%     \label{fig:exampleSingleHunk}
% \end{figure}

\begin{table}[!ht]
	\centering
	\caption{Frequently observed generic patches.}
	\resizebox{1\linewidth}{!}
	{
	\begin{threeparttable}
		\begin{tabular}{l|r|l}
\toprule
          \multicolumn{2}{c|}{Frequency }                     & generic patch \\
\midrule
 \makecell[l]{Hunk \\ Function \\ File \\ Patch \\ Project} & \makecell[r]{202 \\ 99 \\ 99 \\ 116 \\ 1} &
\begin{lstlisting}[backgroundcolor = \color{white},numbers=none,aboveskip=1pt,belowskip=1pt]
@block_content_42_0@
identifier I4, I0;
expression E1, E2, E3;
@@
- struct resource *I0; 
  ...
- I0 = platform_get_resource(E1, IORESOURCE_MEM, E2); 
- E3->I4 = devm_ioremap_resource(&E1->dev, I0); 
+ E3->I4 = devm_platform_ioremap_resource(E1, E2); 
\end{lstlisting} \\
 \bottomrule
   \makecell[l]{Hunk \\ Function \\ File \\ Patch \\ Project} & \makecell[r]{178 \\ 149 \\ 83 \\ 48 \\ 14} &
\begin{lstlisting}[backgroundcolor = \color{white},numbers=none,aboveskip=1pt,belowskip=1pt]
@expr_stmt_4_32@
expression E0;
@@
- return (E0); 
+ return E0; 

\end{lstlisting} \\
   \bottomrule
%   \makecell[l]{Hunk \\ Function \\ File \\ Patch \\ Project} & \makecell[r]{177 \\ 89 \\ 81 \\ 147 \\ 1} &       \lstinputlisting{patternListings/block_content_31_0.cocci} \\
%   \bottomrule
   \makecell[l]{Hunk \\ Function \\ File \\ Patch \\ Project} & \makecell[r]{100 \\ 50 \\ 32 \\ 4 \\ 1} &
\begin{lstlisting}[backgroundcolor = \color{white},numbers=none,aboveskip=1pt,belowskip=1pt]
@block_content_12_25@
expression E0;
@@
- free(E0); 
- E0 = NULL; 
+ FREE_AND_NULL(E0); 

\end{lstlisting}\\
%   \bottomrule
%   \makecell[l]{Hunk \\ Function \\ File \\ Patch \\ Project} &  \makecell[r]{98 \\ 48 \\ 48 \\ 76 \\ 1} &       \lstinputlisting{patternListings/block_content_35_1.cocci2}\\
%   \bottomrule
%   \makecell[l]{Hunk \\ Function \\ File \\ Patch \\ Project} &  \makecell[r]{96 \\ 94 \\ 96 \\ 72 \\ 6} &       \lstinputlisting{patternListings/decl_7_0.cocci1-cocci_patches}\\
%   \bottomrule
%   \makecell[l]{Hunk \\ Function \\ File \\ Patch \\ Project} &  \makecell[r]{88 \\ 22 \\ 22 \\ 4 \\ 1} &       \lstinputlisting{patternListings/expr_stmt_17_1.cocci0}\\
    \bottomrule
   \makecell[l]{Hunk \\ Function \\ File \\ Patch \\ Project} &  \makecell[r]{84 \\ 21 \\ 19 \\ 24 \\ 3} &
\begin{lstlisting}[backgroundcolor = \color{white},numbers=none,aboveskip=1pt,belowskip=1pt]
@block_content_18_3@
expression E2, E1, E0;
@@
- memory_region_init_ram(E0, NULL, E1, E2, &error_abort); 
- vmstate_register_ram_global(E0); 
+ memory_region_allocate_system_memory(E0, NULL, E1, E2); 
\end{lstlisting}\\
%   \bottomrule
%   \makecell[l]{Hunk \\ Function \\ File \\ Patch \\ Project} &  \makecell[r]{82 \\ 36 \\ 36 \\ 3 \\ 2} &       \lstinputlisting{patternListings/if_stmt_11_16.cocci0}\\
   \bottomrule
   \makecell[l]{Hunk \\ Function \\ File \\ Patch \\ Project} &  \makecell[r]{78 \\ 37 \\ 66 \\ 7 \\ 3} &
\begin{lstlisting}[backgroundcolor = \color{white},numbers=none,aboveskip=1pt,belowskip=1pt]
@if_stmt_8_8@
binary operator B1 = {== ,&& };
expression E0, E2;
@@
- if (E0 B1 E2)  
+ if (E2 B1 E0)  
  {
  ...
  }

\end{lstlisting}\\
   
%   \makecell[l]{Hunk \\ Function \\ File \\ Patch \\ Project} &  76 &       \lstinputlisting{patternListings/while_6_1.cocci-cocci_patches}\\
%   \makecell[l]{Hunk \\ Function \\ File \\ Patch \\ Project} &  76 &       \lstinputlisting{patternListings/block_content_38_1.cocci0}\\
%   \makecell[l]{Hunk \\ Function \\ File \\ Patch \\ Project} &  74 &       \lstinputlisting{patternListings/expr_stmt_9_6.cocci0}\\
%   \makecell[l]{Hunk \\ Function \\ File \\ Patch \\ Project} &  74 &       \lstinputlisting{patternListings/decl_stmt_8_3.cocci4}\\
%   \makecell[l]{Hunk \\ Function \\ File \\ Patch \\ Project} &  70 &       \lstinputlisting{patternListings/expr_stmt_12_0.cocci11-cocci7}\\
%   \makecell[l]{Hunk \\ Function \\ File \\ Patch \\ Project} &  65 &       \lstinputlisting{patternListings/decl_13_0.cocci0-cocci_patches}\\
%   \makecell[l]{Hunk \\ Function \\ File \\ Patch \\ Project} &  64 &       \lstinputlisting{patternListings/expr_stmt_15_35.cocci}\\
%   \makecell[l]{Hunk \\ Function \\ File \\ Patch \\ Project} &  62 &       \lstinputlisting{patternListings/return_1_1.cocci0}\\
%   \makecell[l]{Hunk \\ Function \\ File \\ Patch \\ Project} &  61 &       \lstinputlisting{patternListings/if_stmt_8_8.cocci0}\\
%   \makecell[l]{Hunk \\ Function \\ File \\ Patch \\ Project} &  60 &       \lstinputlisting{patternListings/block_content_18_1.cocci}\\
\bottomrule
\end{tabular}

	\end{threeparttable}
	}
	\label{tab:patterns}
\end{table}

%\jk{Anil, what do you want to say in this paragraph. It is detect by Coccinelle. ok and What? this means what?}
We discover that two generic patches (generic patches \#1 and \#3 in Table~\ref{tab:patterns})  have been generated from patches that were actually automatically generate to automate some evolutions at large scale across Linux: the relevant commit logs even mention the Coccinelle tool being used. 

 The second generic patch (id \texttt{expr\_stmt\_4\_32}), is spread among 14 projects (as we can see in the Frequency column of Table~\ref{tab:patterns}) and the associated commits are often described with "Fix coding style" (indeed, the generic patch simply removes brackets).
 The generic patch \texttt{block\_content\_18\_3}  is  inferred  from  3  different  projects. This generic patch fixes   a   memory   mapping   issue.
 Finally, the generic patch \texttt{if\_stmt\_8\_8} switches the order of the expressions in the condition of the \texttt{if statement}, to alter the control flow. A corresponding commit log summarizes this behaviour as \textit{"Put CONFIG* first in if(). This may fix build failures with EAC3 disabled and is more consistent"}.

% \subsection{Pattern Semantinc} \ak{}
To conclude this RQ, we check from which repositories the generic patches have been inferred. Table~\ref{tab:projects}  lists the Top-10 projects which contributed to the pattern inference. We note that all of these projects have  large code histories. Overall, from the 351 repositories used to mine the clusters and infer the generic patterns, 301 contributed to pattern inference. It is possible that the remaining 50 projects do not contribute because of the filtering constraints imposed in our implementation choices, or simply because the do not contain enough recurrent code change context from we generic patches can  be inferred.

\begin{table}[h]
	\centering
	\caption{Top-10  projects contributed to pattern  inference.}
	\resizebox{.8\linewidth}{!}
	{
	\begin{threeparttable}
		\begin{tabular}{l|r|r|r|r|r}
\toprule
        projects    &  freebsd & linux & qemu & wireshark & FFmpeg \\
\midrule
        occurrences & 11812    & 11419 & 9997 & 9337      &  7187   \\
\toprule
\bottomrule      
         projects    & php-src & xqemu & vlc & panda &gtk \\
\midrule
          occurrences &7090    & 6288  & 5249& 4740  & 4325\\
% \midrule
%         freebsd &      11812 \\
%           linux &      11419 \\
%           qemu &       9997 \\
%       wireshark &       9337 \\
%          FFmpeg &       7187 \\
%         php-src &       7090 \\
%           xqemu &       6288 \\
%             vlc &       5249 \\
%           panda &       4740 \\
%             gtk &       4325 \\
            
%           ompi &       4140 \\
%             git &       2694 \\
%         openbsd &       2392 \\
%           prrte &       2190 \\
%           gpdb &       1867 \\
%         cpython &       1776 \\
%       codeflaws &       1714 \\
%       gstreamer &       1605 \\
%         openssl &       1544 \\
%           curl &       1470 \\
%   riscv-openocd &       1122 \\
%         openocd &        990 \\
%  NetworkManager &        980 \\
%       kamailio &        956 \\
%           gpac &        914 \\
\bottomrule
\end{tabular}

% \begin{tabular}{lr}
% \toprule
%         project &  occurence \\
% \midrule
%         freebsd &      11812 \\
%           linux &      11419 \\
%           qemu &       9997 \\
%       wireshark &       9337 \\
%          FFmpeg &       7187 \\
%         php-src &       7090 \\
%           xqemu &       6288 \\
%             vlc &       5249 \\
%           panda &       4740 \\
%             gtk &       4325 \\
%           ompi &       4140 \\
%             git &       2694 \\
%         openbsd &       2392 \\
%           prrte &       2190 \\
%           gpdb &       1867 \\
%         cpython &       1776 \\
%       codeflaws &       1714 \\
%       gstreamer &       1605 \\
%         openssl &       1544 \\
%           curl &       1470 \\
%   riscv-openocd &       1122 \\
%         openocd &        990 \\
%  NetworkManager &        980 \\
%       kamailio &        956 \\
%           gpac &        914 \\
% \bottomrule
% \end{tabular}

	\end{threeparttable}
	}
	\label{tab:projects}
\end{table}

\note{ {\bf RQ-1.2}: \inferrer successfully yields generic patches for a large number of clusters: some generic patches are summarized patterns of changes that spread across several projects.}

\subsection{Generic Patch traceability}
We investigate potential relationships between the distributions of code change locations and the performance of pattern inference, in order to estimate the adequate locations for optimizing the search of generic patches.

\begin{figure}[h]
	\centering
	\includegraphics[width=.8\linewidth]{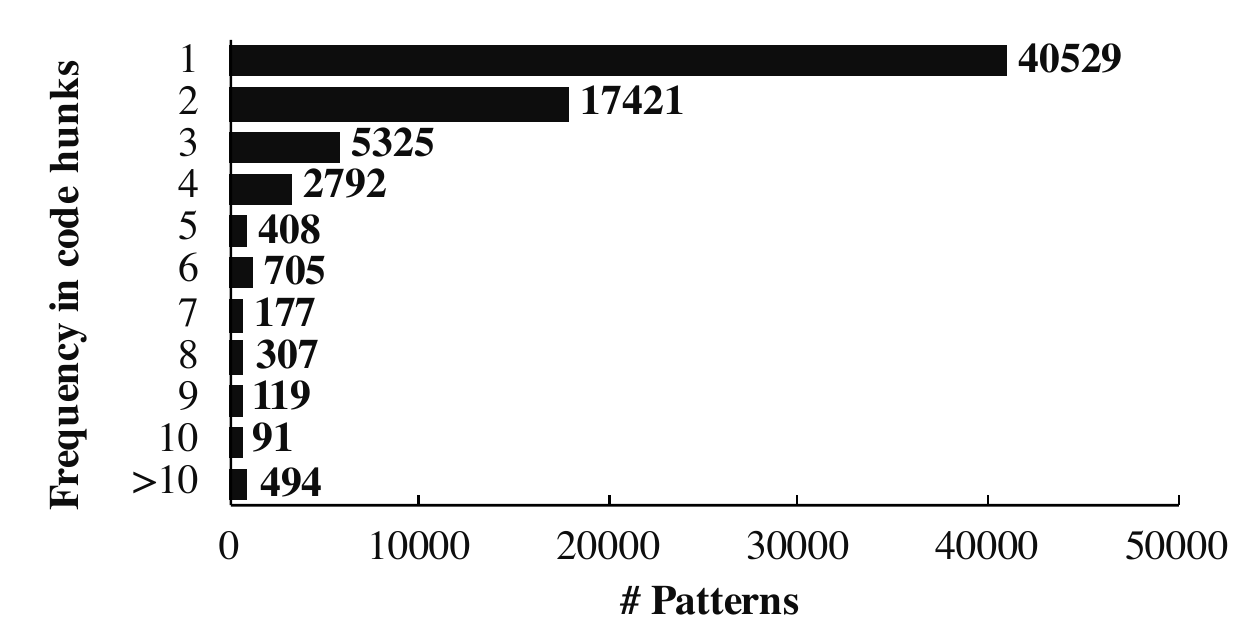}
	\caption{Distribution of the generic patches in hunks.}
	\label{fig:fHunks}
\end{figure}

Generic patches are inferred from hunks in a cluster. Note that, in a cluster, when the code context and change operation of several hunks are syntactically identical we consider them as a single same hunk in this research question. 
Figure~\ref{fig:fHunks} shows the distribution of the generic patches in terms the number of hunks that were used to infer them. Overall, from the \num{68\,368} inferred generic patches, \num{40529} ($\sim$60\%) is inferred from a single hunk.  

On the one hand, we recall that, \miner regroups patches that are similar in terms of code context (AST), and repair action, but we do not consider the similarity among tokens. On the other hand, our current implementation of \inferrer\footnote{which is based on the algorithm of \spinfer~\cite{serrano2020spinfer}} considers also the similarity of the tokens involved in the change to track a pattern: if the tokens involved in the set of patches can be abstracted in a single metavariable then a single transformation rule can be formed. Otherwise, the generic patch will include multiple transformation rules, each including specific tokens (e.g., specific method names).
Figure~\ref{fig:exampleSingleHunk} illustrates a concrete example of such case. For both of the code examples, \miner produces the same AST rich edit script (cf. representation in Fig.~\ref{fig:fixminerPattern}) leading them to be placed in  the same cluster. However, since they differ in terms of the tokens used (different method names and different parameters), \inferrer creates two distinct transformation rules.

\begin{figure}[!ht]
    \centering
    
\begin{lstlisting}[stringstyle=\scriptsize\ttfamily,columns=fullflexible,basicstyle=\scriptsize\ttfamily,aboveskip=2pt,belowskip=2pt]
-       scanf("%s", str);
+       gets(str);
\end{lstlisting}
\begin{lstlisting}[stringstyle=\scriptsize\ttfamily,columns=fullflexible,basicstyle=\scriptsize\ttfamily,aboveskip=2pt,belowskip=2pt]
-       error_setg_errno("%s: stat failed", fname);
+       error_setg_file_open(fname);

\end{lstlisting}
    \caption{An example of patches sharing the same cluster but having distinct generic patches .}
    \label{fig:exampleSingleHunk}
\end{figure}

Note that distributions of generic patches in terms of number of functions and functions also follow the same long tail shape: 
for example, we observed that $\sim$85\% (={58\;176 }/{68\;368}) of the patterns are inferred from a single function. We postulate that the distribution of locations can be used as an heuristics for prioritizing pattern selection in program (cf. RQ-4 for more insights).

\note{{\bf RQ-2}: Generic patches present a long tail distribution in terms of the number of code locations that were involved in their inference. \toolname further provides traceability links to diagnose the code changes set that share similar transformations leading to a fix pattern.}

\subsection{Repairability}
%RQ-3: What is the repairability performance of \toolname?

We assess whether the inferred generic patches can be used to automate generation of patches for real bugs. 
More specifically, 
 we perform two program repair experiments by using the generic patches generated by \inferrer as the main  input  ingredients. 
 Introclass and Codeflaws are leveraged for benchmarking.

Table~\ref{tab:introSota} illustrates the comparative results in terms of numbers of plausible patches (i.e., that make the program pass all the test cases) for the black-box and white-box test suites. Among the selected 764 defects in Introclass, \toolname can generate plausible patches for 186 defects using the black-box test suite and 261 plausible patches using the white-box test suite. Overall, we generate plausible patches for 288 defects of Introclass when both scenarios are combined. We compare the repair performance of \toolname against 3 state-of-the-art APR tools which have been evaluated. 
With the white-box test suite, \toolname ranks second in terms of number of generated plausible patches, and third with the black-box scenario. 
It is noteworthy that \toolname fixes significantly more bugs than other APR tools in some specific projects such as $checksum$, $grade$, and  $syllables$.

% \begin{table}[!ht]
% 	\centering
% 	\scriptsize
% 	\caption{Number of defects repaired plausibly by \toolname on Introclass WB (white-box) and BB(black-box) defects.}
% 	\resizebox{1\linewidth}{!}
% 	{
% 	\begin{threeparttable}
% 		\input{tables/intro}
% 	\end{threeparttable}
% 	}
% 	\label{tab:intro}
% \end{table}

\begin{table}[h]
	\centering
	\caption{Number of Introclass bugs fixed by APR tools.}
	\resizebox{.9\linewidth}{!}
	{
	\begin{threeparttable}

\begin{tabular}{l|rr|rr|rr|rr}
\toprule
          &  \multicolumn{2}{c|}{\toolname} &  \multicolumn{2}{c|}{GenProg} &  \multicolumn{2}{c|}{TrpAutoRepair} &  \multicolumn{2}{c}{AE}\\
  Project &   WB & BB &  WB & BB &  WB & BB    &  WB & BB\\
\midrule
  checksum &  23 & 23 & 3  & 8   &  1   &  0   & 1	&  0	\\
    digits &  37 & 8  & 99 & 30  &  46  &  19  & 50	&  17	\\
     grade &  12 & 8  & 3  & 2   &  2   &  2   & 2	&  2	\\
    median &  44 & 27 & 63 & 108 &  36  &  93  & 16	&  58	\\
  smallest &  75 & 56 & 118& 120 &  118 &  119 & 92	&  71	\\
 syllables &  70 & 64 & 6  & 19  &  9   &  14  & 5	&  11	\\
\bottomrule
Total    & 261 & 186 & 292 & 287 &  212& 247 & 166 & 159
\end{tabular}
						{\footnotesize $^\dagger$ The data about GenProg~\cite{le2011genprog}, TrpAutoRepair~\cite{qi2013efficient} and AE~\cite{weimer2013leveraging}
				are extracted from the experimental results reported by Le Goues~{\em et~al.}~\cite{le2015manybugs}}
	\end{threeparttable}
	}
	\label{tab:introSota}
\end{table}

Table~\ref{tab:blackPatterns} lists example generic patches relevant for fixing Introclass defects. The first generic patch is an example of fix pattern that matches several locations: it substitutes the $C$ standard library function \texttt{scanf()} with \texttt{gets()}. 
According to documentation \texttt{scanf()} reads input until it encounters whitespace, newline or End Of File (EOF), whereas \texttt{gets()} reads input until it encounters newline or End Of File (EOF). We notice that  \emph{gets()} does not stop reading input when it encounters whitespace, but instead it takes whitespace as a string, avoiding bugs.
%Most of the bugs (83) are affected with this whitespace issue.\ak{i didn't verify manually to see if the fix is correct but they pass all the test cases when replaced with this} 
The other listed patterns are mostly related to control logic (i.e., wrong operator usage, boundary checks etc.) in \texttt{if} and \texttt{for} statements. 

\begin{table}[h]
	\centering
	\caption{Selected generic patches fixing Introclass defects.}
	\resizebox{\linewidth}{!}
	{
	\begin{threeparttable}
		\begin{tabular}{l|r||l|r}
\toprule
                                       pattern &  \#defects & pattern &  \#defects \\
\midrule
\begin{lstlisting}[backgroundcolor = \color{white},numbers=none]
@@
expression E0;
@@
- scanf("%s", E0); 
+ gets(E0); 
\end{lstlisting} &     83 &
\begin{lstlisting}[backgroundcolor = \color{white},numbers=none]
@@
expression E0, E1;
@@
- if (E0 ||  E1)  
- {
  ...
- }

\end{lstlisting} &     14 \\
                  \hline
\begin{lstlisting}[backgroundcolor = \color{white},numbers=none]
@@
expression E1;
expression E0;
@@
- if (E0 && E1)  
+ if (E1)  
  {
  ...
  }
- else
+ else
  {
  ...
  }
\end{lstlisting} &     45 
                           &
\begin{lstlisting}[backgroundcolor = \color{white},numbers=none]
@@
expression E3;
expression E2;
expression E1;
expression E0;
@@
- if (E0 < E1 && E2 > E3)  
+ if (E0 <= E1 && E2 >= E3)  
  {
  ...
  }
\end{lstlisting} &     10 \\
                        \hline
\begin{lstlisting}[backgroundcolor = \color{white},numbers=none]
@@
expression E0, E1, E2;
@@
- for(E0 = E1;E0 < E2;E0++)  
+ for(E0 = E1;E0 <= E2;E0++)  
  {
  ...
  }
\end{lstlisting} &      7 &
\begin{lstlisting}[backgroundcolor = \color{white},numbers=none]
@@
expression E2;
expression E0;
binary operator B1 = {< ,>= };
@@
- if (E0 B1 E2)  
+ if (E0 > E2)  
  {
  ...
  }
\end{lstlisting} &      6 \\
%                          \lstinputlisting{patternListings/black/if_stmt_11_71.cocci} &      5 \\
%                       \lstinputlisting{patternListings/black/if_14_157.cocci-cocci7} &      3 \\
%                         \lstinputlisting{patternListings/black/if_stmt_23_89.cocci1} &      2 \\
%                       \lstinputlisting{patternListings/black/if_4_0.cocci25-cocci1} &      2 \\
%                   \lstinputlisting{patternListings/black/block_content_13_46.cocci2} &      2 \\
%                   \lstinputlisting{patternListings/black/block_19_63.cocci0-cocci7} &      1 \\
%                         \lstinputlisting{patternListings/black/if_5_2.cocci7-cocci1} &      1 \\
%                   \lstinputlisting{patternListings/black/block_content_87_25.cocci1} &      1 \\
%                          \lstinputlisting{patternListings/black/if_stmt_10_94.cocci} &      1 \\
%                   \lstinputlisting{patternListings/black/block_content_25_19.cocci1} &      1 \\
%  \lstinputlisting{patternListings/black/if_stmt_6_2.cocci0-cocci2020-07-20_18-02-34} &      1 \\
%                           \lstinputlisting{patternListings/black/if_stmt_8_83.cocci} &      1 \\
\bottomrule
\end{tabular}

	\end{threeparttable}
	}
	\label{tab:blackPatterns}
\end{table}

\begin{table}[h]
	\centering
	\caption{Number of Codeflaws bugs fixed by APR tools.}
	\resizebox{1\linewidth}{!}
	{
	\begin{threeparttable}
			\begin{tabular}{l|>{\columncolor[gray]{0.8}}c|c|c|c|c|c}
	\toprule
	 & { \toolname} & { Angelix} & { Prophet} & { SPR} & { GenProg} & { CoCoNuT}  \\
	\hline
	Total   & { 20/83} &  318/591 & 301/839  &  283/783 &  [255-369]/1423  & 423/716 \\

% 			P(\%) & {\bf 81.3} & 36.2 & 18.5 & 4.5  & 17.7 & 14.3 & 26.1  & 78.3  & 33.3  & 63.4  & 29.0 & 73.1  & {\bf 84.0}  & 60.7 \\
	\bottomrule 
\end{tabular}

				{\footnotesize $^\dagger$ In each column, we provide $x/y$ numbers: $x$ is the number of correctly fixed bugs; $y$ is the number of bugs for which a plausible patch is generated by the APR tool. The data about Angelix~\cite{mechtaev2016angelix}, Prophet~\cite{long2016automatic}, SPR~\cite{long2015staged}, GenProg~\cite{le2011genprog}
				are extracted from the experimental results reported by CoCoNuT~\cite{lutellier2020coconut}.}
	\end{threeparttable}
	}
	\label{tab:codeflawsSota}
\end{table}

Our second experiment is performed on the Codeflaws benchmark. 
For this experiment, we limit the number of generic patches to Top-\num{10000} based on their frequency in code hunks. \toolname can generate plausible patches for 83 defects using the test-suite I.
%, which is given to repair tools for generating repair. 
20 of those 83 defects have been validated to be correctly fixed using the test-suite II. 
We compare the repair performance of \toolname against 5 state-of-the-art APR tools as illustrated in Table~\ref{tab:codeflawsSota}. 
There is an important performance gap between \toolname and other state-of-the-art APR tools. We postulate that the limitation on the number of generic patches had a significant negative impact. Finding a good balance between efficiency (i.e., the search space must not explode by considering all possibilities) and effectiveness can be considered as an engineering detail. We discuss this in the following research question.

\note{{\bf RQ-3}: Note that we do not seek to outperform existing APR tools with our prototype implementation building on the generic patch specifications. Instead, we propose baseline performance for future research in template-based program repair that uses the proposed unified representation of fix patterns. Nevertheless, we note that the baseline is competitive with some state of the art on IntroClass benchmark. }

% https://www.cs.purdue.edu/homes/lintan/publications/coconut-issta20.pdf
% https://arxiv.org/pdf/2005.11040.pdf

% \begin{table}[!ht]
% 	\centering
% 	\caption{Patterns patching introclass with blackbox tests.}
% 	\resizebox{0.5\linewidth}{!}
% 	{
% 	\begin{threeparttable}
% 		\input{tables/whitePatterns}
% 	\end{threeparttable}
% 	}
% 	\label{tab:blackPatterns}
% \end{table}

\subsection{Efficiency } 

We assess efficiency of repair in terms of Number of Patch Candidates (NPC) generated before the first plausible patch is found. NPC represent the invalid patches that an APR tool has consumed resources to test. NPC score has been advocated as a less biased metric of performance compared to execution time~\cite{liu2020efficiency,chen2017contract,ghanbari2019practical}. Our evaluation on the IntroClass benchmark distinguishes two categories:

%The efficiency metric is computed by summing the number of patches in each category:
%$$NPC =NPC_{nonsensical} + NPC_{in-plausible} + NPC_{valid}$$
%
%where: 
\begin{enumerate}[leftmargin=*]
    \item {\bf Nonsensical patches} are patches which cannot even make the patched buggy program successfully compile~\cite{kim2013automatic,monperrus2014critical}.
    \item {\bf In-plausible patches} are patches which let the patched buggy program successfully compile, but fail to pass some test cases in the available test suite.
    % \item {\bf Plausible patch}. Such a patch makes the patched program successfully pass all test cases in the test suites~\cite{qi2015analysis}. A {\em correct patch} is a plausible patch that fixes the bug~\cite{qi2015analysis} (i.e., is not simply overfitted to the test suite). {\em Correctness} is generally determined manually.
\end{enumerate}

% In practice, $NPC_{valid}==1$ since the generation of patches is halted as soon as the first valid patch is found. In this study, since we aim to investigate the repair efficiency, we focus on bugs for which the repair attempts were successfully concluded. Thus, our experimental data do not mention the cases where many patch candidates are generated but none of them was valid. We leave this investigation as a future study.

\begin{figure}[h]
	\centering
	\includegraphics[width=.8\linewidth]{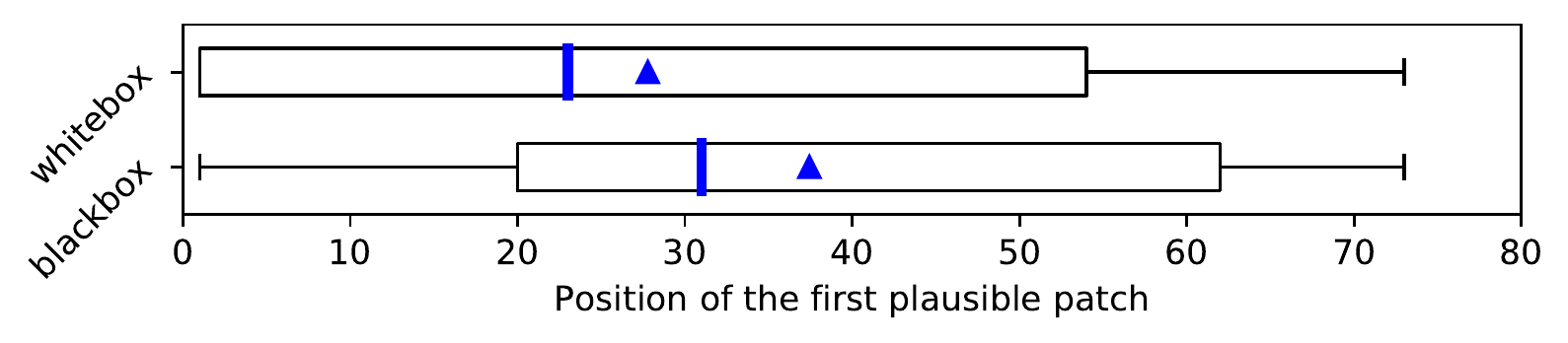}
	\caption{NPC of sensical patches.}
	\label{fig:introPlausiblePos}
\end{figure}

%We compute all following metrics by relying on the Introclass benchmark.  
Figure~\ref{fig:introPlausiblePos} shows the distribution of the position of the first plausible patch when considering all sensical patches (i.e., patches that let the program compile). 
The median of the position  for the black-box scenario is 23, and 31 for the white-box scenario. These represent NPC scores when considering only in-plausible patches.
% This means that, for instance for the black-box scenario, on average, the first 22 patches are in-plausible patches and the 23rd is the first plausible patch.  

\begin{figure}[h]
	\centering
	\includegraphics[width=.8\linewidth]{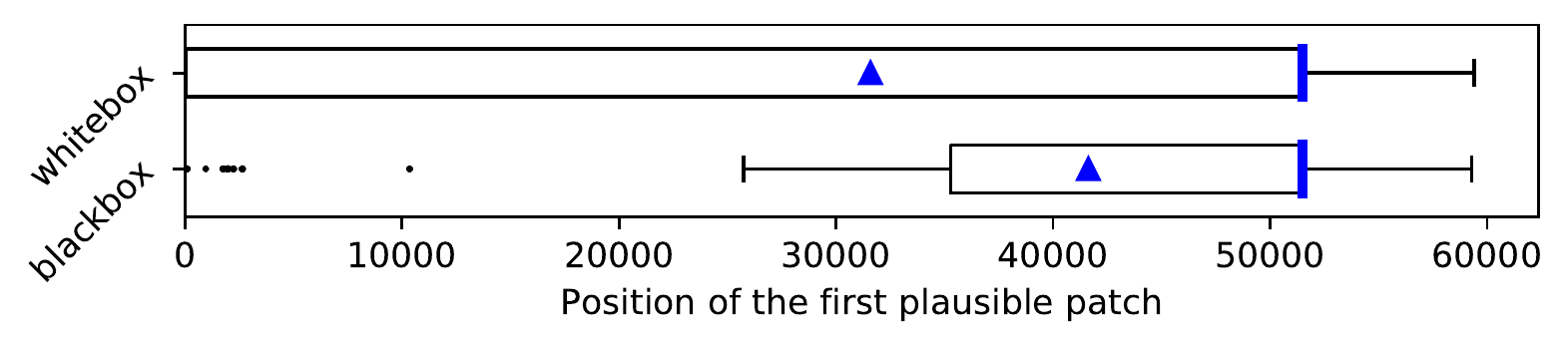}
	\caption{NPC of all patches.}
	\label{fig:introPlausiblePosNonsense}
\end{figure}

Figure~\ref{fig:introPlausiblePosNonsense} shows the distribution of the position of the first plausible patch counting all the patches, i.e., including nonsensical patches). We note that the mean value of NPC in the black-box scenario is \num{41\,626} and in the white-box scenario  \num{31\,587}. 
%Such high values suggest that most of the generic patches that are generating a plausible patch for Introclass benchmark are not frequently observed among the code hunks.
%as during the patch generation the patterns having a higher frequency in code hunks are selected first. 
Such high values indicate that the baseline tool applies first some generic patches that lead to nonsensical patches.
Recall that, we select generic  patches to apply first based  on  their  frequency  in code hunks in \toolname.  
This result therefore suggests that other selection strategies could improve the overall results, and are thus worth to be explored extensively as future work.
%This selection results generating several nonsensical patches before finding the correct generic patch to produce the plausible. 

We investigate a first simple strategy of generic patch selection for repair based on its recurrence in the mining dataset measured in terms of functions, files, patches, projects whose hunks contributed to the pattern abstraction. 
We experiment with all five cases and focus exclusively on white-box scenario. 
%In these experiments, we try 5 different strategies to select generic patches to apply first.

\begin{figure}[h]
	\centering
	\includegraphics[width=1\linewidth]{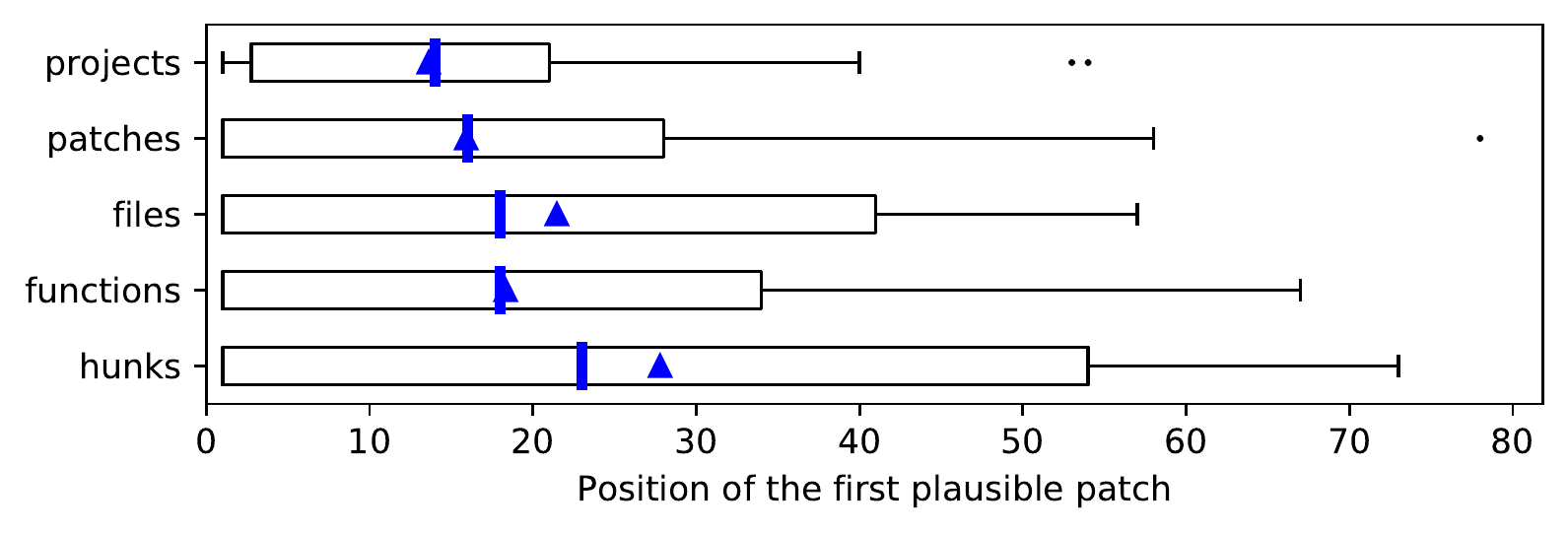}
	\caption{NPC of sensical patches for various selection strategies .}
	\label{fig:whitePatternOrder}
\end{figure}

Figure~\ref{fig:whitePatternOrder} shows the corresponding distribution NPCs excluding the nonsensical patches. 
The strategy where the selection is driven by the frequencies of the generic patches among the projects yields the best results: when we prioritize generic patches inferred from a large number of projects, the NPC score is lower.
% We note that this strategy sharply increases the average NPC.

% we postulate that the  distribution  of  locations  can  be  used  as  an  heuristics  for prioritizing  pattern  selection  in  program  

\begin{figure}[h]
	\centering
	\includegraphics[width=1\linewidth]{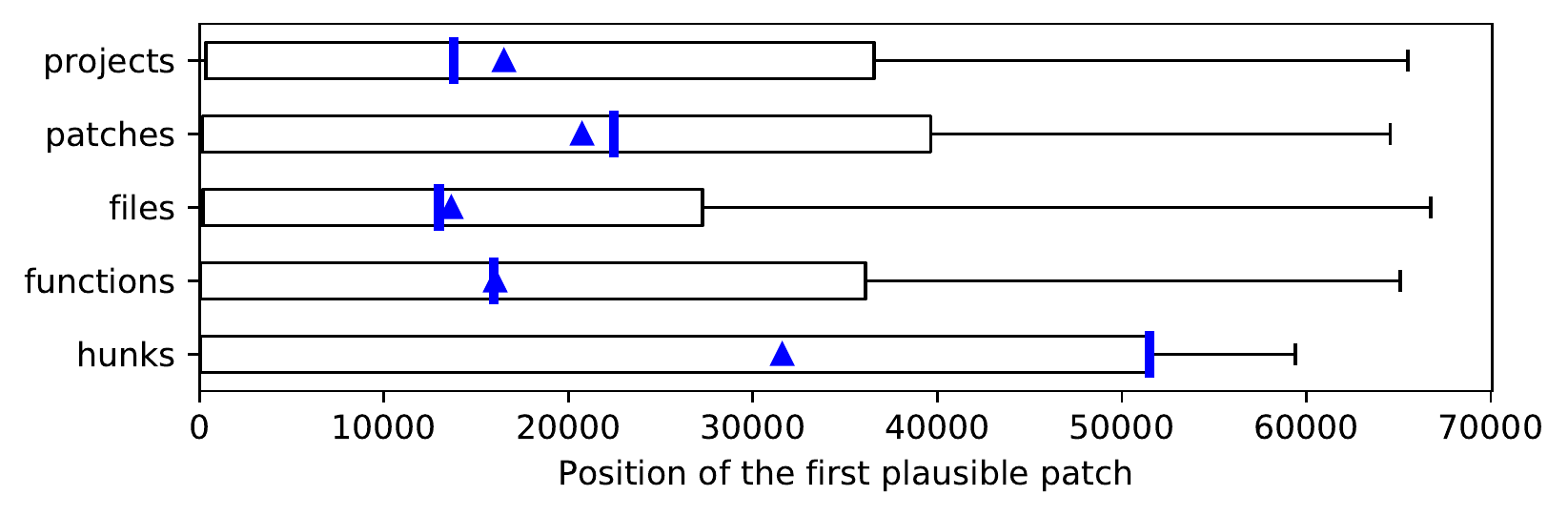}
	\caption{NPC of all patches for various selection strategies.}
	\label{fig:whitePatternOrderNonsense}
\end{figure}

Figure~\ref{fig:whitePatternOrderNonsense} illustrates the NPC considering all patches. Prioritizing generic patches that have been inferred from a large number of projects or a large number of files leads to a significant reduction of NPC by $\sim$48\% (from 31\;587 to 16\;501 when ordered by projects) and 
$\sim$57\% (from 31\;587 to 13\;645 when ordered by files).

%These results suggest that it is necessary to find a good  balance to search space must not explode by considering all possibilities.

\note{{\bf RQ-4}: We note that the efficiency of \toolname could be improved by better prioritizing  generic  patches. We show that because of the traceability in pattern inference, we are able to leverage frequency information to improve the efficiency score by halving the NPC.}

\section{Conclusion}
\label{sec.conclusion}
We have presented \toolname, an open framework for template-based program repair where we build on the concept of generic patch to define a unified  representation/notation for specifying fix patterns (aka templates). We show that generic patches are powerful for expressing fix  patterns in a transparent and  flexible way. \toolname thus offers means, with a baseline, to measure and assess repair new contributions in template-based program repair (e.g., pattern inference, heuristics of candidate search, etc.). 
 We evaluate the repair performance of a prototype implementation on the IntroClass and CodeFlaws benchmarks and we show that our baseline provides comparable performance to the  state  of the art.
We open source \toolname's code and release all data of this study to facilitate replication and encourage further research in this direction: {\bf \url{https://github.com/FlexiRepair}}

\balance
\bibliographystyle{IEEEtran} 
\bibliography{bib/references,bib/referencesFixMiner}

\begin{thebibliography}{10}
\providecommand{\url}[1]{#1}
\csname url@samestyle\endcsname
\providecommand{\newblock}{\relax}
\providecommand{\bibinfo}[2]{#2}
\providecommand{\BIBentrySTDinterwordspacing}{\spaceskip=0pt\relax}
\providecommand{\BIBentryALTinterwordstretchfactor}{4}
\providecommand{\BIBentryALTinterwordspacing}{\spaceskip=\fontdimen2\font plus
\BIBentryALTinterwordstretchfactor\fontdimen3\font minus
  \fontdimen4\font\relax}
\providecommand{\BIBforeignlanguage}[2]{{%
\expandafter\ifx\csname l@#1\endcsname\relax
\typeout{** WARNING: IEEEtran.bst: No hyphenation pattern has been}%
\typeout{** loaded for the language `#1'. Using the pattern for}%
\typeout{** the default language instead.}%
\else
\language=\csname l@#1\endcsname
\fi
#2}}
\providecommand{\BIBdecl}{\relax}
\BIBdecl

\bibitem{liu2020efficiency}
K.~Liu, S.~Wang, A.~Koyuncu, K.~Kim, T.~F. Bissyand{\'e}, D.~Kim, P.~Wu,
  J.~Klein, X.~Mao, and Y.~L. Traon, ``On the efficiency of test suite based
  program repair: A systematic assessment of 16 automated repair systems for
  java programs,'' in \emph{Proceedings of the 42nd International Conference on
  Software Engineering}.\hskip 1em plus 0.5em minus 0.4em\relax ACM, 2020, pp.
  615--627.

\bibitem{bader2019getafix}
J.~Bader, A.~Scott, M.~Pradel, and S.~Chandra, ``Getafix: Learning to fix bugs
  automatically,'' \emph{Proceedings of the ACM on Programming Languages},
  vol.~3, no. OOPSLA, pp. 1--27, 2019.

\bibitem{marginean2019sapfix}
A.~Marginean, J.~Bader, S.~Chandra, M.~Harman, Y.~Jia, K.~Mao, A.~Mols, and
  A.~Scott, ``Sapfix: Automated end-to-end repair at scale,'' in \emph{2019
  IEEE/ACM 41st International Conference on Software Engineering: Software
  Engineering in Practice (ICSE-SEIP)}.\hskip 1em plus 0.5em minus 0.4em\relax
  IEEE, 2019, pp. 269--278.

\bibitem{padioleau2008documenting}
Y.~Padioleau, J.~Lawall, R.~R. Hansen, and G.~Muller, ``Documenting and
  automating collateral evolutions in linux device drivers,'' \emph{Acm sigops
  operating systems review}, vol.~42, no.~4, pp. 247--260, 2008.

\bibitem{lawall2018coccinelle}
J.~Lawall and G.~Muller, ``Coccinelle: 10 years of automated evolution in the
  linux kernel,'' in \emph{2018 {USENIX} Annual Technical Conference}, 2018,
  pp. 601--614.

\bibitem{clever}
M.~{Nayrolles} and A.~{Hamou-Lhadj}, ``Clever: Combining code metrics with
  clone detection for just-in-time fault prevention and resolution in large
  industrial projects,'' in \emph{2018 IEEE/ACM 15th International Conference
  on Mining Software Repositories (MSR)}, 2018, pp. 153--164.

\bibitem{liu2019tbar}
K.~Liu, A.~Koyuncu, D.~Kim, and T.~F. Bissyand{\'e}, ``{TBar}: Revisiting
  template-based automated program repair,'' in \emph{Proceedings of the 28th
  ACM SIGSOFT International Symposium on Software Testing and Analysis}.\hskip
  1em plus 0.5em minus 0.4em\relax ACM, 2019, pp. 31--42.

\bibitem{koyuncu2020fixminer}
A.~Koyuncu, K.~Liu, T.~F. Bissyand{\'e}, D.~Kim, J.~Klein, M.~Monperrus, and
  Y.~Le~Traon, ``Fixminer: Mining relevant fix patterns for automated program
  repair,'' \emph{Empirical Software Engineering}, pp. 1--45, 2020.

\bibitem{ueda2020devreplay}
Y.~Ueda, T.~Ishio, A.~Ihara, and K.~Matsumoto, ``Devreplay: Automatic repair
  with editable fix pattern,'' \emph{arXiv preprint arXiv:2005.11040}, 2020.

\bibitem{koyuncu2019ifixr}
A.~Koyuncu, K.~Liu, T.~F. Bissyand{\'e}, D.~Kim, M.~Monperrus, J.~Klein, and
  Y.~Le~Traon, ``{iFixR}: Bug report driven program repair,'' in
  \emph{Proceedings of the 27the ACM Joint European Software Engineering
  Conference and Symposium on the Foundations of Software Engineering}.\hskip
  1em plus 0.5em minus 0.4em\relax ACM, 2019.

\bibitem{le2019automated}
C.~Le~Goues, M.~Pradel, and A.~Roychoudhury, ``Automated program repair,''
  \emph{Commun. ACM}, 2019.

\bibitem{brunel2009foundation}
J.~Brunel, D.~Doligez, R.~R. Hansen, J.~L. Lawall, and G.~Muller, ``A
  {{Foundation}} for {{Flow}}-based {{Program Matching}}: {{Using Temporal
  Logic}} and {{Model Checking}},'' in \emph{Proceedings of the 36th {{Annual
  ACM SIGPLAN}}-{{SIGACT Symposium}} on {{Principles}} of {{Programming
  Languages}}}, ser. {{POPL}} '09.\hskip 1em plus 0.5em minus 0.4em\relax {New
  York, NY, USA}: {ACM}, 2009, pp. 114--126.

\bibitem{monperrus2018automatic}
M.~Monperrus, ``Automatic software repair: A bibliography,'' \emph{ACM
  Computing Surveys}, vol.~51, no.~1, pp. 17:1--17:24, 2018.

\bibitem{jiang2018shaping}
J.~Jiang, Y.~Xiong, H.~Zhang, Q.~Gao, and X.~Chen, ``Shaping program repair
  space with existing patches and similar code,'' in \emph{Proceedings of the
  27th ACM SIGSOFT International Symposium on Software Testing and
  Analysis}.\hskip 1em plus 0.5em minus 0.4em\relax ACM, 2018, pp. 298--309.

\bibitem{kim2013automatic}
D.~Kim, J.~Nam, J.~Song, and S.~Kim, ``Automatic patch generation learned from
  human-written patches,'' in \emph{Proceedings of the 35th International
  Conference on Software Engineering}.\hskip 1em plus 0.5em minus 0.4em\relax
  IEEE, 2013, pp. 802--811.

\bibitem{westley2009automatically}
W.~Weimer, T.~Nguyen, C.~{Le Goues}, and S.~Forrest, ``Automatically finding
  patches using genetic programming,'' in \emph{Proceedings of the 31st
  International Conference on Software Engineering, May 16-24,}.\hskip 1em plus
  0.5em minus 0.4em\relax Vancouver, Canada: {IEEE}, 2009, pp. 364--374.

\bibitem{mechtaev2016angelix}
S.~Mechtaev, J.~Yi, and A.~Roychoudhury, ``Angelix: Scalable multiline program
  patch synthesis via symbolic analysis,'' in \emph{Proceedings of the 38th
  International Conference on Software Engineering}.\hskip 1em plus 0.5em minus
  0.4em\relax ACM, 2016, pp. 691--701.

\bibitem{nguyen2013semfix}
H.~D.~T. Nguyen, D.~Qi, A.~Roychoudhury, and S.~Chandra, ``Semfix: Program
  repair via semantic analysis,'' in \emph{Proceedings of the 35th
  International Conference on Software Engineering}.\hskip 1em plus 0.5em minus
  0.4em\relax IEEE, 2013, pp. 772--781.

\bibitem{xiong2017precise}
Y.~Xiong, J.~Wang, R.~Yan, J.~Zhang, S.~Han, G.~Huang, and L.~Zhang, ``Precise
  condition synthesis for program repair,'' in \emph{Proceedings of the 39th
  IEEE/ACM International Conference on Software Engineering}.\hskip 1em plus
  0.5em minus 0.4em\relax IEEE, 2017, pp. 416--426.

\bibitem{gupta2017deepfix}
R.~Gupta, S.~Pal, A.~Kanade, and S.~Shevade, ``Deepfix: Fixing common c
  language errors by deep learning,'' in \emph{Proceedings of the 31st AAAI
  Conference on Artificial Intelligence}.\hskip 1em plus 0.5em minus
  0.4em\relax {AAAI}, 2017, pp. 1345--1351.

\bibitem{long2017automatic}
F.~Long, P.~Amidon, and M.~Rinard, ``Automatic inference of code transforms for
  patch generation,'' in \emph{Proceedings of the 11th Joint Meeting on
  Foundations of Software Engineering}.\hskip 1em plus 0.5em minus 0.4em\relax
  ACM, 2017, pp. 727--739.

\bibitem{long2016automatic}
F.~Long and M.~Rinard, ``Automatic patch generation by learning correct code,''
  in \emph{Proceedings of the 43rd Annual {ACM} {SIGPLAN-SIGACT} Symposium on
  Principles of Programming Languages}, vol.~51, no.~1.\hskip 1em plus 0.5em
  minus 0.4em\relax ACM, 2016, pp. 298--312.

\bibitem{urli2018design}
S.~Urli, Z.~Yu, L.~Seinturier, and M.~Monperrus, ``How to design a program
  repair bot?: insights from the repairnator project,'' in \emph{Proceedings of
  the 40th International Conference on Software Engineering: Software
  Engineering in Practice}.\hskip 1em plus 0.5em minus 0.4em\relax ACM, 2018,
  pp. 95--104.

\bibitem{scott2019getafix}
A.~Scott, J.~Bader, and S.~Chandra, ``Getafix: Learning to fix bugs
  automatically,'' \emph{arXiv preprint arXiv:1902.06111}, 2019.

\bibitem{claire2012genprog}
C.~{Le Goues}, T.~Nguyen, S.~Forrest, and W.~Weimer, ``Genprog: {A} generic
  method for automatic software repair,'' \emph{{IEEE} Trans. Software Eng.},
  vol.~38, no.~1, pp. 54--72, 2012.

\bibitem{liu2018mining}
X.~Liu and H.~Zhong, ``Mining stackoverflow for program repair,'' in
  \emph{Proceedings of the 25th IEEE International Conference on Software
  Analysis, Evolution and Reengineering}.\hskip 1em plus 0.5em minus
  0.4em\relax IEEE, 2018, pp. 118--129.

\bibitem{hua2018towards}
J.~Hua, M.~Zhang, K.~Wang, and S.~Khurshid, ``Towards practical program repair
  with on-demand candidate generation,'' in \emph{Proceedings of the 40th
  International Conference on Software Engineering}.\hskip 1em plus 0.5em minus
  0.4em\relax ACM, 2018, pp. 12--23.

\bibitem{saha2017elixir}
R.~K. Saha, Y.~Lyu, H.~Yoshida, and M.~R. Prasad, ``Elixir: Effective
  object-oriented program repair,'' in \emph{Proceedings of the 32nd IEEE/ACM
  International Conference on Automated Software Engineering}.\hskip 1em plus
  0.5em minus 0.4em\relax IEEE, 2017, pp. 648--659.

\bibitem{durieux2017dynamic}
T.~Durieux, B.~Cornu, L.~Seinturier, and M.~Monperrus, ``Dynamic patch
  generation for null pointer exceptions using metaprogramming,'' in
  \emph{Proceedings of the 24th International Conference on Software Analysis,
  Evolution and Reengineering}.\hskip 1em plus 0.5em minus 0.4em\relax IEEE,
  2017, pp. 349--358.

\bibitem{martinez2018ultra}
M.~Martinez and M.~Monperrus, ``Ultra-large repair search space with
  automatically mined templates: the cardumen mode of astor,'' in
  \emph{International Symposium on Search Based Software Engineering}.\hskip
  1em plus 0.5em minus 0.4em\relax Springer, 2018, pp. 65--86.

\bibitem{liu2019avatar}
K.~Liu, A.~Koyuncu, D.~Kim, and T.~F. Bissyand{\'e}, ``Avatar: Fixing semantic
  bugs with fix patterns of static analysis violations,'' in \emph{Proceedings
  of the 26th IEEE International Conference on Software Analysis, Evolution and
  Reengineering}.\hskip 1em plus 0.5em minus 0.4em\relax IEEE, 2019, pp.
  456--467.

\bibitem{liu2019you}
K.~Liu, A.~Koyuncu, T.~F. Bissyand{\'e}, D.~Kim, J.~Klein, and Y.~L. Traon,
  ``You cannot fix what you cannot find! an investigation of fault localization
  bias in benchmarking automated program repair systems,'' in \emph{Proceedings
  of the 12th IEEE International Conference on Software Testing, Verification
  and Validation}.\hskip 1em plus 0.5em minus 0.4em\relax IEEE, 2019.

\bibitem{liu2018mining2}
K.~Liu, D.~Kim, T.~F. Bissyand{\'e}, S.~Yoo, and Y.~Le~Traon, ``Mining fix
  patterns for findbugs violations,'' \emph{IEEE Transactions on Software
  Engineering}, 2018.

\bibitem{wen2018context}
M.~Wen, J.~Chen, R.~Wu, D.~Hao, and S.-C. Cheung, ``Context-aware patch
  generation for better automated program repair,'' in \emph{Proceedings of the
  40th International Conference on Software Engineering}.\hskip 1em plus 0.5em
  minus 0.4em\relax ACM, 2018, pp. 1--11.

\bibitem{fluri2008discovering}
B.~Fluri, E.~Giger, and H.~C. Gall, ``Discovering patterns of change types,''
  in \emph{Proceedings of the 23rd IEEE/ACM International Conference on
  Automated Software Engineering}.\hskip 1em plus 0.5em minus 0.4em\relax
  L'Aquila, Italy: IEEE, 2008, pp. 463--466.

\bibitem{martinez2015mining}
M.~Martinez and M.~Monperrus, ``Mining software repair models for reasoning on
  the search space of automated program fixing,'' \emph{Empirical Software
  Engineering}, vol.~20, no.~1, pp. 176--205, 2015.

\bibitem{fluri2006classifying}
B.~Fluri and H.~C. Gall, ``Classifying change types for qualifying change
  couplings,'' in \emph{Program Comprehension, 2006. ICPC 2006. 14th IEEE
  International Conference on}.\hskip 1em plus 0.5em minus 0.4em\relax IEEE,
  2006, pp. 35--45.

\bibitem{pan2009toward}
K.~Pan, S.~Kim, and E.~J. Whitehead, ``Toward an understanding of bug fix
  patterns,'' \emph{Empirical Software Engineering}, vol.~14, no.~3, pp.
  286--315, 2009.

\bibitem{kim2009discovering}
M.~Kim and D.~Notkin, ``Discovering and representing systematic code changes,''
  in \emph{Proceedings of the 31st International Conference on Software
  Engineering}.\hskip 1em plus 0.5em minus 0.4em\relax IEEE Computer Society,
  2009, pp. 309--319.

\bibitem{kim2006memories}
S.~Kim, K.~Pan, and E.~Whitehead~Jr, ``Memories of bug fixes,'' in
  \emph{Proceedings of the 14th ACM SIGSOFT international symposium on
  Foundations of software engineering}.\hskip 1em plus 0.5em minus 0.4em\relax
  ACM, 2006, pp. 35--45.

\bibitem{molderez2017mining}
T.~Molderez, R.~Stevens, and C.~De~Roover, ``Mining change histories for
  unknown systematic edits,'' in \emph{Proceedings of the 14th International
  Conference on Mining Software Repositories}.\hskip 1em plus 0.5em minus
  0.4em\relax IEEE Press, 2017, pp. 248--256.

\bibitem{yue2017characterization}
R.~Yue, N.~Meng, and Q.~Wang, ``A characterization study of repeated bug
  fixes,'' in \emph{Software Maintenance and Evolution (ICSME), 2017 IEEE
  International Conference on}.\hskip 1em plus 0.5em minus 0.4em\relax IEEE,
  2017, pp. 422--432.

\bibitem{nguyen2010recurring}
T.~T. Nguyen, H.~A. Nguyen, N.~H. Pham, J.~Al-Kofahi, and T.~N. Nguyen,
  ``Recurring bug fixes in object-oriented programs,'' in \emph{Software
  Engineering, 2010 ACM/IEEE 32nd International Conference on}, vol.~1.\hskip
  1em plus 0.5em minus 0.4em\relax IEEE, 2010, pp. 315--324.

\bibitem{liu2018closer}
K.~Liu, D.~Kim, A.~Koyuncu, L.~Li, T.~F. Bissyand{\'e}, and Y.~Le~Traon, ``A
  closer look at real-world patches,'' in \emph{2018 IEEE International
  Conference on Software Maintenance and Evolution (ICSME)}.\hskip 1em plus
  0.5em minus 0.4em\relax IEEE, 2018, pp. 275--286.

\bibitem{livshits2005dynamine}
B.~Livshits and T.~Zimmermann, ``{{DynaMine}}: {{Finding Common Error
  Patterns}} by {{Mining Software Revision Histories}},'' in \emph{Proceedings
  of the 10th {{European Software Engineering Conference Held Jointly}} with
  13th {{ACM SIGSOFT International Symposium}} on {{Foundations}} of {{Software
  Engineering}}}, ser. {{ESEC}}/{{FSE}}-13.\hskip 1em plus 0.5em minus
  0.4em\relax {New York, NY, USA}: {ACM}, 2005, pp. 296--305.

\bibitem{hanam2016discovering}
Q.~Hanam, F.~S. d.~M. Brito, and A.~Mesbah, ``Discovering bug patterns in
  javascript,'' in \emph{Proceedings of the 2016 24th ACM SIGSOFT International
  Symposium on Foundations of Software Engineering}.\hskip 1em plus 0.5em minus
  0.4em\relax ACM, 2016, pp. 144--156.

\bibitem{chawathe1996change}
S.~S. Chawathe, A.~Rajaraman, H.~{Garcia-Molina}, and J.~Widom, ``Change
  {{Detection}} in {{Hierarchically Structured Information}},'' in
  \emph{Proceedings of the 1996 {{ACM SIGMOD International Conference}} on
  {{Management}} of {{Data}}}, ser. {{SIGMOD}} '96.\hskip 1em plus 0.5em minus
  0.4em\relax {New York, NY, USA}: {ACM}, 1996, pp. 493--504.

\bibitem{andersen2010generic}
J.~Andersen and J.~L. Lawall, ``Generic patch inference,'' \emph{Automated
  software engineering}, vol.~17, no.~2, pp. 119--148, 2010.

\bibitem{andersen2012semantic}
J.~Andersen, A.~C. Nguyen, D.~Lo, J.~L. Lawall, and S.-C. Khoo, ``Semantic
  patch inference,'' in \emph{Automated Software Engineering (ASE), 2012
  Proceedings of the 27th IEEE/ACM International Conference on}.\hskip 1em plus
  0.5em minus 0.4em\relax IEEE, 2012, pp. 382--385.

\bibitem{serrano2020spinfer}
\BIBentryALTinterwordspacing
L.~Serrano, V.-A. Nguyen, F.~Thung, L.~Jiang, D.~Lo, J.~Lawall, and G.~Muller,
  ``{SPINFER}: Inferring semantic patches for the linux kernel,'' in \emph{2020
  {USENIX} Annual Technical Conference ({USENIX} {ATC} 20)}.\hskip 1em plus
  0.5em minus 0.4em\relax {USENIX} Association, Jul. 2020, pp. 235--248.
  [Online]. Available:
  \url{https://www.usenix.org/conference/atc20/presentation/serrano}
\BIBentrySTDinterwordspacing

\bibitem{durieux2019analysis}
T.~Durieux, R.~Abreu, M.~Monperrus, T.~F. Bissyand{\'e}, and L.~Cruz, ``An
  analysis of 35+ million jobs of travis ci,'' in \emph{2019 IEEE International
  Conference on Software Maintenance and Evolution (ICSME)}.\hskip 1em plus
  0.5em minus 0.4em\relax IEEE, 2019, pp. 291--295.

\bibitem{le2015manybugs}
C.~Le~Goues, N.~Holtschulte, E.~K. Smith, Y.~Brun, P.~Devanbu, S.~Forrest, and
  W.~Weimer, ``The manybugs and introclass benchmarks for automated repair of c
  programs,'' \emph{IEEE Transactions on Software Engineering}, vol.~41,
  no.~12, pp. 1236--1256, 2015.

\bibitem{tan2017codeflaws}
S.~H. Tan, J.~Yi, S.~Mechtaev, A.~Roychoudhury \emph{et~al.}, ``Codeflaws: a
  programming competition benchmark for evaluating automated program repair
  tools,'' in \emph{Proceedings of the 39th International Conference on
  Software Engineering Companion}.\hskip 1em plus 0.5em minus 0.4em\relax IEEE
  Press, 2017, pp. 180--182.

\bibitem{cadar2008klee}
C.~Cadar, D.~Dunbar, D.~R. Engler \emph{et~al.}, ``Klee: unassisted and
  automatic generation of high-coverage tests for complex systems programs.''
  in \emph{OSDI}, vol.~8, 2008, pp. 209--224.

\bibitem{le2011genprog}
C.~Le~Goues, T.~Nguyen, S.~Forrest, and W.~Weimer, ``Genprog: A generic method
  for automatic software repair,'' \emph{Ieee transactions on software
  engineering}, vol.~38, no.~1, pp. 54--72, 2011.

\bibitem{qi2013efficient}
Y.~Qi, X.~Mao, and Y.~Lei, ``Efficient automated program repair through
  fault-recorded testing prioritization,'' in \emph{2013 IEEE International
  Conference on Software Maintenance}.\hskip 1em plus 0.5em minus 0.4em\relax
  IEEE, 2013, pp. 180--189.

\bibitem{weimer2013leveraging}
W.~Weimer, Z.~P. Fry, and S.~Forrest, ``Leveraging program equivalence for
  adaptive program repair: Models and first results,'' in \emph{Proceedings of
  the 28th IEEE/ACM International Conference on Automated Software
  Engineering}.\hskip 1em plus 0.5em minus 0.4em\relax IEEE, 2013, pp.
  356--366.

\bibitem{long2015staged}
F.~Long and M.~Rinard, ``Staged program repair with condition synthesis,'' in
  \emph{Proceedings of the 10th Joint Meeting on Foundations of Software
  Engineering}.\hskip 1em plus 0.5em minus 0.4em\relax ACM, 2015, pp. 166--178.

\bibitem{lutellier2020coconut}
T.~Lutellier, H.~V. Pham, L.~Pang, Y.~Li, M.~Wei, and L.~Tan, ``Coconut:
  combining context-aware neural translation models using ensemble for program
  repair,'' in \emph{Proceedings of the 29th ACM SIGSOFT International
  Symposium on Software Testing and Analysis}, 2020, pp. 101--114.

\bibitem{chen2017contract}
L.~Chen, Y.~Pei, and C.~A. Furia, ``Contract-based program repair without the
  contracts,'' in \emph{Proceedings of the 32nd IEEE/ACM International
  Conference on Automated Software Engineering}.\hskip 1em plus 0.5em minus
  0.4em\relax IEEE, 2017, pp. 637--647.

\bibitem{ghanbari2019practical}
A.~Ghanbari, S.~Benton, and L.~Zhang, ``Practical program repair via bytecode
  mutation,'' in \emph{Proceedings of the 28th ACM SIGSOFT International
  Symposium on Software Testing and Analysis}.\hskip 1em plus 0.5em minus
  0.4em\relax ACM, 2019, pp. 19--30.

\bibitem{monperrus2014critical}
M.~Monperrus, ``A critical review of automatic patch generation learned from
  human-written patches: essay on the problem statement and the evaluation of
  automatic software repair,'' in \emph{Proceedings of the 36th International
  Conference on Software Engineering}.\hskip 1em plus 0.5em minus 0.4em\relax
  ACM, 2014, pp. 234--242.

\end{thebibliography}

\end{document}